\documentclass[aps,prd,showpacs,notitlepage,nofootinbib,superscriptaddress,floatfix,twocolumn,showkeys]{revtex4-1}

\usepackage{graphicx}
\usepackage{amsmath}
\usepackage{amsfonts}
\usepackage{mathtools}

\usepackage{physics}
\usepackage{amssymb}
\usepackage{bm}
\usepackage{graphicx}
\usepackage{amssymb}
\usepackage{amsmath}
\usepackage{bm}
\usepackage{latexsym}
\usepackage{epsfig}
\usepackage{psfrag}
\usepackage{color}
\usepackage{comment}
\usepackage{amsmath}
\usepackage{orcidlink}

\usepackage{hyperref}

\def\l{\left}
\def\r{\right}
\def\f{\frac}
\def\e{\mathrm{e}}

\def\Ni{N_{\mathrm{i}}}
\def\Nic{N_{\mathrm{ic}}}

\def\d{{\mathrm{d}}}
\def\Mpl{M_{_{\mathrm{Pl}}}}
\def\nn{\nonumber} 
\def\ps{{\mathcal P}_{_{\mathrm{S}}}}
\def\As{A_{_{\mathrm{S}}}}
\def\ns{n_{_{\mathrm{S}}}}

\def\cG{\mathcal{G}}

\def\tanh{\rm{tanh}}

\allowdisplaybreaks

\begin{document}


\title{Reconstructing inflationary features on large scales using genetic algorithm}
\author{Alipriyo Hoory}
\email{E-mail: alipriyo@physics.iitm.ac.in}
\affiliation{Centre for Strings, Gravitation and Cosmology,
Department of Physics, Indian Institute of Technology Madras, 
Chennai~600036, India}
\author{Dhiraj Kumar Hazra}
\email{E-mail: dhiraj@imsc.res.in}
\affiliation{The Institute of Mathematical Sciences, HBNI, CIT Campus, Chennai 
600113, India}
\affiliation{Homi Bhabha National Institute, Training School Complex, Anushakti 
Nagar, Mumbai~400085, India}
\affiliation{INAF/OAS Bologna, Osservatorio di Astrofisica e Scienza dello Spazio, 
Area della ricerca CNR-INAF, via Gobetti 101, I-40129 Bologna, Italy}
\author{L.~Sriramkumar}
\email{E-mail: sriram@physics.iitm.ac.in}
\affiliation{Centre for Strings, Gravitation and Cosmology,
Department of Physics, Indian Institute of Technology Madras, 
Chennai~600036, India}
\begin{abstract}
As is well known, inflationary models involving a single, slowly rolling, scalar 
field predict a smooth, nearly scale-invariant and featureless scalar power spectrum,
which is remarkably consistent with the observed anisotropies in the cosmic 
microwave background (CMB) and the distribution of the large-scale structure.
However, a variety of model-dependent as well as model-independent approaches 
suggest that certain localized features in the scalar power spectrum can lead 
to a significantly better fit to the CMB data.
In this work, we focus on three types of such features and examine whether these 
features can be generated in inflationary scenarios driven by a single, canonical
scalar field.
We consider a slowly rolling baseline model that is described by a specific 
time-dependence of the first slow roll parameter and we generate the desired 
features in the power spectrum through suitable modifications to the functional 
form of the slow roll parameter.
To systematically reconstruct the desired features in the scalar power spectrum 
(or, equivalently, the modifications in the behavior of the first slow roll 
parameter) that are consistent with the data, we implement a machine learning 
pipeline based on the genetic algorithm~(GA).
Assuming the standard values for the background~$\Lambda$CDM model (arrived at for 
a nearly scale-invariant primordial scalar power spectrum), we apply our method to 
the Planck 2018 CMB data, and show that the reconstructed features improve the fit
to the observed angular power spectra by~$\Delta \chi^2 \lesssim -10$. 
Moreover, we find that GA points to other sets of background parameters and primordial
features, which lead to a similar level of improvement in the fit to the data.
Such alternative sets of background parameters and scalar power spectra offer 
possible pathways to alleviate existing cosmological tensions. 
Our approach provides effective single-field inflationary dynamics to generate 
features that are supported by the data.
\end{abstract}
\maketitle


\section{Introduction}\label{sec:introduction}

Inflation is the most promising framework for describing the primordial universe 
(see, for example, the reviews~\cite{Mukhanov:1990me,Martin:2003bt,Martin:2004um,
Bassett:2005xm,Sriramkumar:2009kg,Baumann:2008bn,Baumann:2009ds,
Sriramkumar:2012mik,Linde:2014nna,Martin:2015dha}). 
It refers to a brief period of accelerated expansion during the early stages of 
the radiation-dominated epoch, and it was originally proposed to overcome the
shortcomings of the standard hot big bang model, such as the horizon and flatness
problems. 
In addition to resolving these issues, inflation naturally generates the primordial
perturbations.
The perturbations originate from the quantum fluctuations associated with the scalar 
fields that drive inflation.
The fluctuations that are present during the initial stages of inflation are stretched
to cosmological scales during inflation.
After inflation, as the universe expands further, the perturbations evolve, leaving 
their imprints as anisotropies in the cosmic microwave background (CMB) and leading
to the large-scale structure (LSS) that we observe around us today. 
The observations of the anisotropies in the CMB and the distribution of the LSS 
suggest that the primordial fluctuations are largely adiabatic and Gaussian in 
nature~\cite{Planck:2018jri, Planck:2019kim}. 
In such a situation, the power spectrum of the primordial scalar (or curvature) 
perturbations quantifies the statistical distribution of the fluctuations, and 
it plays a central role in connecting the theory to the observations. 
The observations point to a nearly scale-invariant spectrum of primordial scalar
perturbations, as is generated in simple models of slow roll inflation
involving a single, canonical scalar field (for a detailed comparison of such 
models with the CMB observations, see Ref.~\cite{Martin:2013tda}).
Over the coming years, ongoing and upcoming missions focused on observing the 
polarization of the CMB as well as surveys of galaxies are expected to provide
further data with significantly higher precision (in this context, see 
Refs.~\cite{LiteBIRD:2023zmo,LiteBIRD:2024wix,Euclid:2024yrr,ACT:2025tim,
SPT-3G:2025bzu,SimonsObservatory:2025wwn}).
There is great hope that these observations can help us gain deeper insights 
into the physics operating in the early universe.

One of the key goals of such precision cosmology is to uncover signatures of
any departures from nearly scale-invariant primordial scalar power spectra.
In other words, the aim is to determine whether the data point to features in 
the primordial scalar power spectrum (for detailed discussions in this regard, 
see, for instance, Refs.~\cite{Chluba:2015bqa,Chen:2016vvw,Chen:2016zuu,
Euclid:2023shr}).
These features can be generated due to non-trivial dynamics during inflation, 
thereby hinting at physics beyond the simple models.
It has been repeatedly argued in the literature that certain features in the 
primordial scalar power spectrum improve the fit to the CMB data (for a partial 
list of such efforts, see Refs.~\cite{Bridle:2003sa,Cline:2003ve,Contaldi:2003zv,
Shafieloo:2003gf,Shafieloo:2006hs,Shafieloo:2007tk,Jain:2009pm,Nicholson:2009pi,
Nicholson:2009zj,Hazra:2010ve,Aich:2011qv,Hazra:2013xva,Hazra:2013eva,Chen:2014cwa,
Hazra:2014jwa,Ragavendra:2020old,Braglia:2021rej,Antony:2022ert,Sohn:2022jsm}). 
While there is yet to be a definitive detection of features in the CMB, the 
Planck data suggest more than a handful of statistically significant outliers---i.e.
departures from the predictions of slow roll inflationary models---across 
large and small angular scales~\cite{Planck:2018jri}.
Such outliers can be potential signatures of primordial features and they
have sparked considerable theoretical and observational interest (for discussions 
in this regard, see, for instance, Refs.~\cite{Chen:2014cwa,Antony:2022ert}). 
In the era of high-precision data, modern computational tools such as machine 
learning have emerged as powerful techniques to model and analyze complex 
cosmological phenomena. 
In particular, machine learning methods are well-suited for exploring 
higher-dimensional parameter spaces and extracting patterns from the 
data. 
In this context, it becomes imperative to investigate whether such tools can be 
employed to lift degeneracies among competing inflationary models and also to 
construct non-trivial scenarios that are significantly more consistent with the 
current observations than the simple models. 

In this work, we shall investigate three types of features in the primordial 
scalar power spectrum.
The first type of feature that we shall consider is a localized burst of 
oscillations in the primordial spectrum.
Such a feature can be generated by introducing damped oscillations with a
Gaussian envelope (DOGE) in the first slow roll parameter.
Recall that, when the inflationary models are compared with the observations,
generically, the background parameters are varied along with the parameters
describing the inflationary models.
The often quoted best-fit values for the background cosmological parameters 
assume that the primordial scalar power spectrum is of a nearly scale-invariant 
form, as is generated in slow roll inflation~\cite{Planck:2018jri}.
When one introduces features in the scalar power spectrum, the values of 
the best-fit parameters of the background cosmological model as well as 
the other derived parameters can shift from the values obtained when the
standard, featureless scalar power spectrum is considered. 
Interestingly, in addition to improving the fit to the Planck data, it has 
been shown that the features generated by DOGE in the first slow roll 
parameter alleviate the so-called~$H_0$ and 
$S_8$~tensions~\cite{Schoneberg:2021qvd,Antony:2022ert}.
(Note that, while $H_0$ is the Hubble parameter today, $S_8$~refers to the 
strength of the fluctuations in the matter density.)
Compared to the conventional nearly scale-invariant scalar power spectrum as 
is generated in slow roll inflation, these features lead to a larger value 
of~$H_0$ and a smaller value of~$S_8$. 
The data from future CMB missions are expected to improve the constraints 
on such features~\cite{LiteBIRD:2023zmo,LiteBIRD:2024wix,Euclid:2024yrr,
ACT:2025tim,SPT-3G:2025bzu,SimonsObservatory:2025wwn}.

The second type of feature that we shall consider is aimed at simultaneously 
fitting the outliers in the CMB data at the low multipoles of $\ell<30$ and
around the high multipoles of $\ell\simeq 750$.
To fit such outliers, an analytical spectrum was initially proposed in the 
so-called classical primordial standard clock (CPSC) model wherein massive 
fields oscillating about the minima induce periodic features in the
scalar power spectrum due to resonance (for the original discussion, see 
Ref.~\cite{Chen:2014cwa}).
More recently, a two-field model has been constructed which generates a 
scalar power spectrum of a similar form and leads to a better fit to the outliers
in the CMB data at the multipoles we indicated above~\cite{Braglia:2021rej}.
We should mention that, currently in the literature, there is no effective
single-field scenario that is able to fit these outliers at the low and high 
multipoles simultaneously.
One of our aims in this work is to construct such a scenario.

The third type of scalar power spectrum with features that we shall consider 
is one which has actually been reconstructed from the CMB data.
In astronomical imaging, the observed image is a convolution of the original
image and the point spread function (PSF) of the imaging system.
(Note that the PSF is the response of an imaging system to a point source.) 
In this context, the so-called Richardson-Lucy (RL) algorithm has been developed 
to iteratively reconstruct the underlying image from the observed image and 
the known PSF~\cite{Richardson:72}. 
In cosmology, as is well known, the angular power spectra of the CMB are 
a convolution of the primordial power spectra and the transfer functions
which depend only on the background (in this context, see the standard 
texts~\cite{Weinberg:2008zzc,Dodelson:2020bqr,Durrer:2020fza}).
Assuming a given background cosmological model, the modified RL (MRL) 
algorithm has been utilized to reconstruct the primordial scalar power spectrum 
from the CMB data~\cite{Shafieloo:2003gf,Shafieloo:2006hs, Shafieloo:2007tk, 
Nicholson:2009pi,Nicholson:2009zj,Hazra:2013eva,Hazra:2013xva,Hazra:2014jwa}.
More recently, there has been an effort to modify the algorithm suitably
to reduce the overfitting of the noise in the data~\cite{Sohn:2022jsm}.
The reconstructed primordial scalar spectrum contains many features and
we should mention that there has been constant efforts in the literature 
to generate one or more of such features from specific inflationary 
models (see, for example, Refs.~\cite{Cline:2003ve,Contaldi:2003zv,
Jain:2009pm,Hazra:2010ve,Aich:2011qv,Chen:2014cwa,Ragavendra:2020old,
Braglia:2021rej,Antony:2022ert}).

In this work, we shall adopt a non-parametric machine learning tool called
the genetic algorithm (GA) to arrive at the best-fit inflationary scalar power 
spectra.
Assuming that inflation is driven by a single, canonical scalar field, we 
shall model the background evolution analytically in terms of the 
time-dependence of the first slow roll parameter.
As we shall see, the GA will help us arrive at effective single-field models 
that result in a good fit to the data.
We shall consider three analytical forms for the first slow roll parameter, 
all of which have comparable levels of significance. 
We shall initially consider the case wherein the first slow roll parameter 
contains one or more DOGEs. 
Thereafter, as a second functional form for the first slow roll parameter, 
we shall consider linear combinations of tangent hyperbolic and sinusoidal 
functions, which lead to spectra similar to the ones generated in the 
CPSC model. 
To mimic the power spectra reconstructed from the MRL algorithm, we shall 
consider an analytical form for the first slow roll parameter that involves
a linear combination of tangent hyperbolic and damped sinusoidal functions. 
After choosing suitable priors for the parameters involved, we shall allow 
GA to find the optimized functional forms of the first slow roll parameter. 
We shall make use of the Planck 2018 data---specifically, the data incorporated
in the likelihood referred to as \texttt{Plik-TTTEEE+lowl+lowE+lensing}---to 
determine the optimized functional forms~\cite{Planck:2019nip}. 
Eventually, we shall also plot the residuals for the best-fit CMB angular power
spectra to illustrate the manner in which the inflationary scalar power spectra 
with features outperform the conventional nearly scale-invariant spectra at 
the different multipoles. 

This manuscript is organized as follows.
In the following section, we shall first describe the manner in which we shall
model the background evolution during inflation in terms of the time-dependence
(to be exact, the dependence on $e$-folds) of the first slow roll parameter. 
We shall introduce the baseline slow roll model and describe the numerical 
module to calculate the inflationary scalar power spectrum. 
Thereafter, we shall discuss in detail the implementation of the GA and the 
pipeline we have developed to introduce features in the scalar power spectrum
(through the modifications to the time-dependence of the first slow roll 
parameter) and compare the models with the data.
In Sec.~\ref{sec:test-run}, we shall present the results of a test run that 
we have carried out to check the validity of our pipeline. 
Specifically, we shall make use of a known scalar power spectrum as a target 
and demonstrate the ability of the GA to reconstruct it effectively. 
In Sec.~\ref{sec:choice-of-function}, we shall discuss the functional forms 
that modify the behavior of the first slow roll parameter to generate specific 
features in the inflationary scalar power spectrum. 
We shall briefly discuss the motivations for considering each modification
and indicate the range of priors we shall choose to work with for the 
parameters involved.
In Sec.~\ref{sec:results}, we shall present the results from the GA and 
discuss the constraints arrived at using the Planck 2018 data for the three
types of features.
We shall also illustrate the resulting CMB angular power spectra computed
using~\texttt{CLASS}~\cite{Lesgourgues:2011re,Blas:2011rf}.
In Sec.~\ref{sec:results-cgb}, we shall work with a different set of 
background parameters and make use of GA to arrive at the functional 
form of the first slow roll parameter that fits the data well.
Such an exercise illustrates the manner in which features in the inflationary
scalar power spectrum can aid in alleviating the~$H_0$ and~$S_8$ tensions.
In Sec.~\ref{sec:potential}, from the best-fit functional forms of the 
first slow roll parameter, we shall numerically reconstruct the
corresponding inflationary potentials.  
Finally, in Sec.~\ref{sec:summary}, after a brief summary of our findings, 
we shall discuss broader the implications of our results as well as possible 
directions for further investigations.

At this stage of our discussion, we should make a few clarifying remarks
concerning the conventions and notations that we shall work with.
We shall work with natural units such that $\hbar=c=1$ and set the reduced 
Planck mass to be $\Mpl=\l(8 \pi G\r)^{-1/2}$.
We shall assume the background to be the spatially flat
Friedmann-Lema\^itre-Robertson-Walker~(FLRW) universe described by the 
following line-element:
\begin{equation}
\d s^2=-\d t^2+a^2(t)\,\d {\bm x}^2
=-a^{2}(\eta) \l(\d \eta^2-\d {\bm x}^2\r),\label{eq:flrw-le}
\end{equation}
where~$t$ and~$\eta$ are the cosmic and conformal time coordinates, 
and~$a$ denotes the scale factor.
Also, an overdot and an overprime shall denote differentiation with 
respect to~$t$ and $\eta$, respectively.
Moreover, $H=\dot{a}/a$ shall denote the Hubble parameter associated
with the FLRW universe.
Further, $N$ shall represent the number of $e$-folds.


\section{Methodology}\label{sec:methodology}

In this section, we shall first discuss the baseline inflationary model 
that serves as the foundation for our analysis.
We shall also briefly outline the numerical calculation of the inflationary 
scalar power spectrum.
Thereafter, we shall describe the GA and the pipeline we construct to 
compare the inflationary scalar power spectrum containing features with 
the CMB data.


\subsection{Baseline model and the scalar power spectrum}\label{subsec:e-sps}

Let us first introduce the baseline inflationary scenario that we shall 
consider.
The purpose of the baseline scenario is to provide a simple, observationally 
consistent model on top of which we can later introduce modifications (to
lead to features in the inflationary scalar power spectrum) and test their 
performance against the CMB data using GA. 

As is well-known, in single-field models of inflation driven by the 
canonical scalar field, the dynamics during inflation can be described 
completely in terms of the slow roll parameters.
In fact, it is sufficient to specify the functional form (i.e. the 
dependence in time) of the first slow roll parameter, as the higher-order 
slow roll parameters can be constructed in terms of the first.
Therefore, we shall begin by considering a particular functional form~(in
terms of $e$-folds $N$) of the first slow roll parameter which is motivated 
by the standard, slow roll inflationary scenario.
Recall that the first slow roll parameter $\epsilon_1$ is defined in terms
of the Hubble parameter $H$ as $\epsilon_1=-\dot{H}/H^2=-\d \ln H/\d N$.
We shall assume that the baseline inflationary model is described by the 
following form of the first slow roll parameter~\cite{Antony:2022ert}:
\begin{equation}
\epsilon_{1\mathrm{bl}}(N)
= \epsilon_{1a}\mathrm{exp}[\epsilon_{2a}(N-N_1)],\label{eq:e1-bl}
\end{equation}
where $(\epsilon_{1a},\epsilon_{2a},N_1)$ are constants.
Later, we shall choose values for these constants to ensure that the 
primary observables related to the inflationary scalar and tensor power 
spectra---viz. the scalar amplitude~$\As$, the scalar spectral index~$\ns$
and the tensor-to-scalar ratio~$r$---are in agreement with the current
cosmological data.

Having discussed the background evolution during inflation, let us turn to 
the perturbations.
Besides solving the horizon problem associated with the hot big bang model,
inflation also provides a mechanism for the origin of the primordial
perturbations. 
The primordial perturbations are expected to originate from the quantum
vacuum during inflation.
In the single-field models of inflation of our interest, the scalar 
perturbations are often described by the so-called curvature perturbation 
(in this regard, see, for instance, the reviews~\cite{Mukhanov:1990me,
Martin:2003bt,Martin:2004um,Bassett:2005xm,Sriramkumar:2009kg,Baumann:2008bn,
Baumann:2009ds,Sriramkumar:2012mik,Linde:2014nna,Martin:2015dha}).
The homogeneity and isotropy of the FLRW universe allows us to decompose 
the curvature perturbation in terms of the Fourier modes functions, say, 
$f_k$, where $k$ denotes the wave number of the perturbation.
It can be shown that the mode functions $f_k$ satisfy the following equation 
of motion~\cite{Mukhanov:1990me,Martin:2003bt,Martin:2004um,Bassett:2005xm,
Sriramkumar:2009kg,Baumann:2008bn,Baumann:2009ds,Sriramkumar:2012mik,
Linde:2014nna,Martin:2015dha}:
\begin{equation}
f_{k}''+2 \frac{z'}{z} f_{k}'+k^{2} f_{k}=0,\label{eq:de-fk}
\end{equation}
where $z=\sqrt{2 \epsilon_1} \Mpl a$, with $\epsilon_1$ being the first slow
roll parameter. 
The power spectrum of the scalar perturbations is the two-point correlation
function of the curvature perturbations in Fourier space.
The scalar power spectrum generated during inflation, say, $\ps(k)$, can be 
expressed in terms of the Fourier mode functions~$f_k$ as follows:
\begin{equation}
\ps(k)=\f{k^3}{2\pi^2} \vert f_k\vert^2.
\end{equation}

The scalar power spectrum generated during inflation leaves its imprints 
on the CMB and the LSS, which are the key observables for testing the
inflationary models. 
Therefore, to connect the dynamics during inflation with the observations, 
we need to compute the scalar power spectrum in the model at hand. 
Evidently, arriving at the scalar power spectrum requires solving the
equation of motion~\eqref{eq:de-fk} governing the mode functions~$f_k$ 
for a given set of initial conditions.
The Fourier mode functions~$f_k$ can be solved analytically in simple 
situations such as slow roll inflation.
In general, when departures from slow roll inflation are involved, one has
to resort to the numerical evaluation of the mode functions and the power
spectrum (in this context, see, for example, Refs.~\cite{Hazra:2012yn,
Ragavendra:2020old}).
It proves to be efficient to solve the differential equation~\eqref{eq:de-fk} 
in terms of the number of $e$-folds~$N$.
If we use $e$-folds as the independent variable, the equation
of motion~\eqref{eq:de-fk} is modified to the form
\begin{equation}
\f{\d^2f_{k}}{\d N^2}
+\l(3- \epsilon_{1}+\epsilon_{2}\r) \f{\d f_{k}}{\d N}
+\f{k^2}{a^2 H^2} f_k=0,\label{eq:de-fk-N}
\end{equation}
where $\epsilon_2=\d \ln \epsilon_1/\d N$ is the second slow roll parameter.
As is well-known, during inflation, the initial conditions on the perturbations
are imposed when the Fourier modes are well inside the Hubble radius, i.e. when
$k \gg a H$. 
Typically, the initial conditions on the Fourier mode functions with the wave 
number~$k$ are imposed at an $e$-fold, say, $N_\mathrm{ic}$, such that $k\simeq 
100\, a H$.
At such a time, one imposes the following initial conditions on the mode
functions~$f_k$ and their time-derivatives~$\d f_k/\d N$:
\begin{subequations}\label{eq:bd-ic}
\begin{align}
f_{k}(\Nic) &=\f{1}{\sqrt{2k}\, z(\Nic)},\\
\f{\d f_{k}}{\d N}\biggl\vert_{\Nic}
&=-\f{1}{\sqrt{2k}  z(\Nic)} 
\biggl[\f{i k}{a(\Nic)H(\Nic)}\nn\\
&\quad+ \f{1}{z(\Nic)}\f{\d z}{\d N}\biggl\vert_{\Nic}\biggr].
\end{align}
\end{subequations}
These conditions are often referred to as the Bunch-Davies initial 
conditions~\cite{Bunch:1977sq,Bunch:1978yq,Bunch:1978yw}.
We should mention that, in these initial conditions, we have omitted an overall
phase factor which does not affect the physical results.
Once these initial conditions are set, the differential equation~\eqref{eq:de-fk-N} 
can be evolved numerically until late times, close to the end of inflation,
to evaluate the scalar power spectrum.

It is the values of the parameters involved and the dynamics of the
inflationary background that determine the amplitude and shape of 
the scalar power spectrum.
Note that the differential equation~\eqref{eq:de-fk-N} governing the 
Fourier mode function~$f_k$ involves the background quantities~$a$, $H$, 
$\epsilon_1$ and $\epsilon_2$.
The scale factor can be expressed in terms of $e$-folds as $a(N)=a_\ast 
\exp (N-N_\ast)$, where $a_\ast$ is fixed by the condition that the pivot 
scale, say, $k_\ast$, leaves the Hubble radius $N_\ast$ $e$-folds before
the end of inflation.
Recall that we shall be describing the dynamics of the inflationary 
background in terms of the functional dependence of the first slow roll
parameter~$\epsilon_1(N)$. 
Given $\epsilon_1(N)$, the second slow roll parameter $\epsilon_2(N)$ 
can be immediately determined.
Also, since $\epsilon_1=\d \ln H/\d N$, given an initial value of 
the Hubble parameter, we can integrate the $\epsilon_1(N)$ we consider 
to obtain~$H(N)$. 
With the background quantities and the initial conditions on~$f_k$ at hand,
we can evolve~$f_k$ using a standard numerical procedure and evaluate the 
inflationary scalar power spectrum.

To compute the scalar power spectrum~$\ps(k)$, we have developed 
a \texttt{Python}-based numerical module that solves the differential 
equation~\eqref{eq:de-fk-N} for a wide range of comoving wavenumbers~$k$. 
Given the functional form of the first slow roll parameter~$\epsilon_1(N)$, 
our code evaluates the background dynamics, interpolates $a(N)$ and $H(N)$, 
and evolves the mode function~$f_k(N)$ in the manner described above.
The Bunch-Davies initial conditions~\eqref{eq:bd-ic} are imposed 
sufficiently deep inside the Hubble radius and the mode functions 
are evolved until their amplitude freezes on super-Hubble scales, 
typically when $k \lesssim 10^{-5}\, a H$. 

In our analysis, we shall choose the values of the three parameters 
in $\epsilon_{1\mathrm{bl}}(N)$, i.e. in the functional form of the 
baseline model [cf. Eq.~\eqref{eq:e1-bl}], to 
be: $(\epsilon_{1a},\epsilon_{2a},N_1)=(1.65 \times 10^{-3},
3.24\times 10^{-2},20)$.
For such values, it can be readily checked that $\epsilon_{1\mathrm{bl}}(N)
\simeq 1$ when $N\simeq 218$.
In other words, inflation proves to be rather long.
Therefore, we shall terminate inflation {\it by hand}\/ after~$72$ 
$e$-folds.
We should clarify that, if needed, an end to inflation can be achieved 
naturally by modifying the form of $\epsilon_{1\mathrm{bl}}(N)$ at 
later times without affecting its behavior during the initial stages.
However, since we are focusing on the inflationary scalar power spectrum 
over large scales, say, over the wave numbers $10^{-5} \lesssim k \lesssim
1\, \mathrm{Mpc}^{-1}$, these details need not be a concern.
We have chosen $N_1$ such that the pivot scale of $k_\ast= 0.05\,
\mathrm{Mpc}^{-1}$ exits the Hubble radius $N_\ast=52$ $e$-folds 
before the end of inflation. 
Recall that, in the slow roll approximation, the scalar amplitude~$\As$,
the scalar spectral index~$\ns$ and the tensor-to-scalar ratio~$r$ are 
given by~\cite{Mukhanov:1990me,Martin:2003bt,Martin:2004um,Bassett:2005xm,
Sriramkumar:2009kg,Baumann:2008bn,Baumann:2009ds,Sriramkumar:2012mik,
Linde:2014nna,Martin:2015dha} 
\begin{subequations}
\begin{align}
\As&=\f{H_\ast^2}{8\pi^2\Mpl^2\epsilon_{1\ast}}
\l[1-2(C+1)\epsilon_{1\ast}-C\epsilon_{2\ast}\r],\\
\ns&=1-2\epsilon_{1\ast}
-\epsilon_{2\ast},\\
r&=16\epsilon_{1\ast},
\end{align}
\end{subequations}
where $C=\gamma_{\mathrm{_{E}}}-2+\ln2\simeq -0.73$, with $\gamma_{\mathrm{_{E}}}$
being the Euler-Mascheroni constant, and the asterisks denote the 
values at the time when the pivot scale leaves the Hubble radius.
For the baseline parameterization~\eqref{eq:e1-bl} of the first slow roll
parameter and the choice of the values of $(\epsilon_{1a},\epsilon_{2a},N_1)$ 
mentioned above, we can analytically determine the scalar spectral index and 
the tensor-to-scalar ratio to be $\ns \simeq 0.964$ and $r \simeq 0.0264$.
We should point out that these values are consistent with the latest data from 
Planck as well as the BICEP-Keck array~\cite{Planck:2018jri,BICEP:2021xfz}.
The scalar amplitude of $\ln (10^{10} \As)=3.038$ at the pivot scale
suggested by the Planck data determines the initial value of the 
Hubble parameter to be $H(N_{\mathrm{i}})/\Mpl=H_{\mathrm{i}}/\Mpl
=1.67\times 10^{-5}$.
We should add that these values determined analytically match the values we 
obtain from the numerical evaluation of the scalar and tensor power spectra.
We should also mention that the value of the goodness of fit~$\chi^2$ 
corresponding to the baseline model is $\chi^2=2774.09$, when the model
is compared with the Planck 2018 data, specifically, when the 
~\texttt{Plik}~likelihood is used along with the \texttt{TTTEEE+lowl+lowE+lensing}
data~\cite{Planck:2019nip}.
In our discussion below, we shall focus on the scalar power spectrum and 
ignore the contributions due to the tensors.


\subsection{Genetic algorithm}

The GA is a stochastic optimization method inspired by natural selection, 
capable of searching large and complex functional spaces without relying 
on gradients~\cite{book1,10.5555/522098}. 
Unlike deterministic optimization schemes, the GA works by evolving a 
population of trial solutions over successive generations, gradually 
improving their fitness according to a well defined evolutionary strategy. 
This makes it particularly effective for non-parametric, analytical 
reconstructions where the underlying functional form is not known in 
advance and the search space is highly non-linear~\cite{Bogdanos:2009ib, 
Nesseris:2012tt,Arjona:2019fwb,Arjona:2020kco,Kamerkar:2022dfu, 
Lodha:2023jru,Medel-Esquivel:2023nov}. 
In this work, we apply the GA to reconstruct the evolution of the first 
slow roll parameter~$\epsilon_1(N)$ directly from cosmological observations. 
We shall suitably modify and utilize the GA module that has been developed 
and used earlier in different cosmological contexts~\cite{Bogdanos:2009ib,
Nesseris:2012tt,Arjona:2019fwb,Arjona:2020kco,Kamerkar:2022dfu}.
The key advantage of this approach is that it allows us to explore a wide
class of candidate inflationary histories without imposing restrictive 
assumptions. 

The GA proceeds in the following steps:  
\begin{itemize}
\item 
\textbf{Initialization:}~A grammar and priors are specified to generate
the initial population of candidate functions. 
The grammar, in general, includes basic functions such as polynomials, 
exponentials, logarithms and trigonometric functions.
These functions are combined with the elementary mathematical operations
of addition, subtraction, multiplication and division to arrive at the 
candidate functions.
From these building blocks, the algorithm randomly generates $N_{\rm pop}
=100$ trial forms for~$\epsilon_1(N)$ in the first generation. 
Evidently, the priors have to be chosen such that the scenario is physically 
viable.
Specifically, in our context, we choose the priors such that~$\epsilon_1(N) 
\ll 1$ during the span of time (i.e. the period of inflation) of our interest.  
\item 
\textbf{Evaluation:}~The inflationary scalar power spectrum is computed for
each candidate~$\epsilon_1(N)$ using the \texttt{Python}-based module in 
the manner we have described in the previous section.
The resulting scalar power spectrum is then passed to the Boltzmann 
code~\texttt{CLASS}, which calculates the CMB angular power 
spectra~$\mathcal{C}_\ell$. 
Thereafter, the~$\mathcal{C}_\ell$'s that have been computed are compared
with the CMB data. 
Specifically, we make use of the Planck likelihood \texttt{Plik} to assign
each candidate $\epsilon_1(N)$ a value for the goodness of fit~$\chi^2$, 
which acts as the fitness score. 
\item 
\textbf{Selection:}~Candidates are ranked by fitness, i.e. by the 
quantity~$\chi^2$. 
We employ the so-called tournament selection scheme, in which four random 
individuals are compared and the best two are selected for 
reproduction~\cite{Bogdanos_2009}. 
This balances stochasticity and evolutionary pressure, avoiding premature 
convergence while steadily improving the population.  
\item 
\textbf{Crossover:}~Selected individuals are recombined to generate offsprings
using a single-point crossover~\cite{GoldbergGA,Michalewicz1996GeneticA}. 
In this implementation, the parameters describing each candidate~$\epsilon_1(N)$ 
is represented as a list and a random crossover point along the parameter list 
determines how the parent information is exchanged. 
The first offspring is created by taking the first segment of the first parent 
up to the crossover point and combining it with the remaining segment from the 
second parent, while the second offspring is constructed from the complementary 
segments. 
Formally, for parent individuals, say, $\text{parent}_0$ and $\text{parent}_1$ 
with length~$L$, a random integer $l_\mathrm{c} \in [1, L-1]$ defines the 
crossover point, producing the following two offsprings:
\begin{subequations}
\begin{align}
\text{offspring}_1 &= \text{parent}_0[:l_\mathrm{c}] 
+ \text{parent}_1[l_\mathrm{c}:],\\
\text{offspring}_2 &= \text{parent}_0[l_\mathrm{c}:]
+ \text{parent}_1[:l_\mathrm{c}].
\end{align}
\end{subequations}
The crossover is applied probabilistically with a crossover rate $p_\mathrm{c} 
= 0.8$, which ensures that the majority of the population participates in 
recombining each generation, thus promoting exploration of the solution space. 
\item 
\textbf{Mutation:}~To maintain genetic diversity and explore the solution 
space effectively, we implement a custom uniform random 
mutation~\cite{GoldbergGA,Michalewicz1996GeneticA}. 
For each individual, a random number between $0$ and $1$ is drawn, and if it is 
smaller than the specified mutation rate (in our case, we choose it to be
$p_\mathrm{m} = 0.4$), the mutation operator is applied to that individual. 
The mutation selects one parameter of the candidate individual at random and 
replaces its value with a new number drawn uniformly from the predefined allowed 
range for that parameter. 
Each parameter has its own range, ensuring that the mutation respects physically 
motivated bounds. 
Formally, for an individual with parameters $(X_0, X_1, \dots)$, the mutated value 
$X'_i$ is given by
\begin{equation}
X'_i \sim \mathrm{Uniform}[\text{range}(X_i)],
\end{equation}
where the index $i$ is chosen randomly for each mutation event. 
This probabilistic application of mutation ensures that roughly half of the 
population is modified each generation, balancing exploration of new regions 
of the parameter space and enabling efficient global search.
\item 
\textbf{Replacement and elitism:}~Offsprings replace the least-fit individuals,
but the best candidates (elites) are always preserved. 
This guarantees monotonic improvement in the fitness score~$\chi^2$ across 
generations.  
\end{itemize}

In summary, the GA offers a distinct advantage over traditional parameter
estimation methods such as the Markov chain Monte Carlo (MCMC) method or 
grid-based searches. 
While MCMC techniques offer a powerful tool for exploring well-defined parameter 
spaces around specific models, they often become computationally expensive and
inefficient when the underlying functional form of the model is unknown or when 
the parameter space is highly non-linear and degenerate. 
In contrast, GA does not rely on predefined parametric forms and instead performs 
a non-parametric functional reconstruction through iterative evolution of the
candidate solutions. 
This flexibility allows GA to efficiently explore a broader functional landscape, 
identify non-trivial features in the inflationary scalar power spectrum, and avoid 
getting trapped in local minima. 
Moreover, since GA is population-based, it naturally incorporates a global search 
strategy rather than a local exploration, making it particularly well-suited for 
uncovering features that would otherwise remain hidden in conventional parameter 
searches.  
Its flexibility allows us to discover functional forms of $\epsilon_1(N )$ that
would be inaccessible to rigid parametric methods, while its evolutionary nature 
ensures the exploration of a broader class of inflationary scenarios.


\subsection{Pipeline}\label{subsec:pipeline}

While the GA controls the optimization strategy, the pipeline determines
the manner in which candidate functions are physically mapped to cosmological 
observables and compared with the data. 
Specifically, the GA supplies trial forms of $\epsilon_1(N)$, and the 
pipeline connects them to predictions for the CMB, thereby enabling 
direct comparison with the observations.  
We have suitably modified the publicly available Genetic Algorithm 
Toolkit to incorporate our custom selection, crossover, and mutation 
techniques within our GA pipeline (in this regard, see 
Refs.~\cite{Bogdanos:2009ib,Nesseris:2012tt,Arjona:2019fwb,Arjona:2020kco,
Kamerkar:2022dfu}).

We shall express the modified first slow roll parameter as follows:
\begin{equation}
\epsilon_{1}(N)
= \epsilon_{1\mathrm{bl}}(N) [1 + F(N)],\label{eq:m-fsrp}
\end{equation}
where $\epsilon_{1\mathrm{bl}}(N)$ is the behavior of the baseline 
first slow roll parameter given by Eq.~\eqref{eq:e1-bl} and $F(N)$ 
is a linear combination of up to four functions that will be 
introduced by the GA to describe the departures from slow roll. 
Such a parameterization leads to deviations from the smooth baseline 
evolution, which generate specific features in the inflationary scalar 
power spectrum.

Within the pipeline, the GA first constructs the functional form of the
first slow roll parameter, which is used to compute the corresponding
inflationary scalar power spectrum.
This is computed by the \texttt{Python}-based module that we described earlier
in Sec.~\ref{subsec:e-sps}. 
The scalar power spectrum is then passed to the Boltzmann solver~\texttt{CLASS} 
to calculate the CMB angular power spectra~\cite{Lesgourgues:2011re,Blas:2011rf}.
Thereafter, the pipeline calculates the corresponding cumulative $\chi^2$ for 
the $TT$, $TE$ and $EE$ CMB angular power spectra.
To do so, we employ the full Planck 2018 data, i.e. we make use of the 
\texttt{Plik}~likelihood along with the \texttt{TTTEEE+lowl+lowE+lensing}
data~\cite{Planck:2019nip}.
In other words, the~\texttt{Plik}~likelihood incorporates temperature and 
polarization measurements at both high and low multipoles as well as the 
effects due to gravitational lensing~\cite{Planck:2019nip}.
Based on the goodness of fit~$\chi^2$, the population of candidate solutions 
is ranked, and the next generation is produced through the selection, crossover, 
and mutation procedures described above. 
The process is iterated over successive generations, allowing the GA 
to progressively improve the fit to the data.

\begin{figure*}[t]
\includegraphics[width=0.75\textwidth]{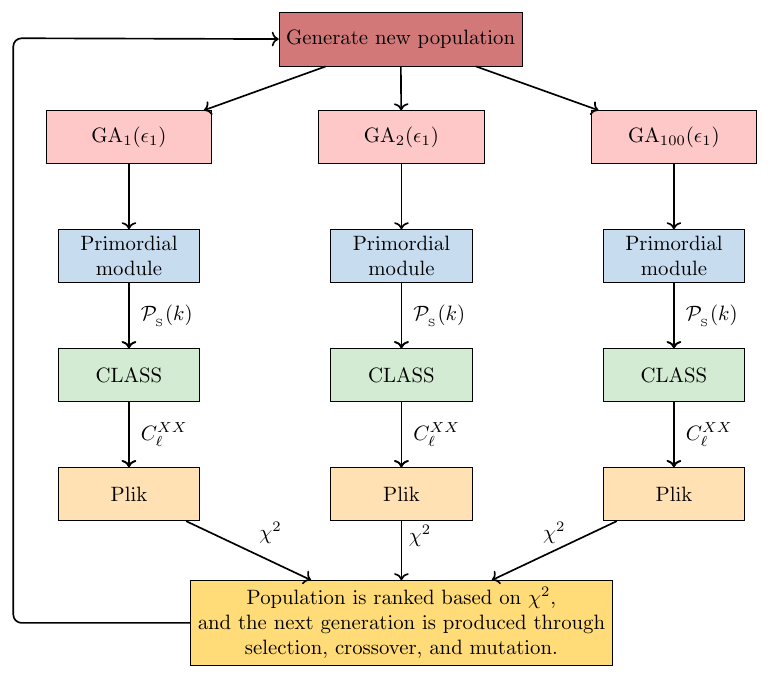}
\caption{A schematic flowchart illustrating our pipeline. 
The GA begins by generating a population of $100$ candidate functions of 
the first slow roll parameter~$\epsilon_1(N)$ over a specific range of 
priors. 
Each function~$\epsilon_1(N)$ is then passed to the primordial module to 
compute the inflationary scalar power spectrum~$\ps(k)$.
The resulting scalar power spectrum provides input to the Boltzmann 
code~\texttt{CLASS} to calculate the angular power spectra $\mathcal{C}_{\ell}^{XX}$
(where $X$ refers to $T$ or $E$) of the CMB. 
These angular power spectra are compared against observational data using 
the Planck likelihood module~\texttt{Plik} to arrive at the log-likelihood, 
i.e. the goodness of fit $\chi^2$, for each model. 
The process is repeated over $400$ times to arrive at the functional form 
of the first slow roll parameter $\epsilon_1(N)$ that fits the data the
best.}\label{GA_flowchart}
\end{figure*}
A schematic overview of the algorithm is shown in Fig.~\ref{GA_flowchart}. 
The evolutionary cycle is repeated for $400$ generations, corresponding 
to a total of $4\times10^{4}$ model evaluations.
The primary computational cost arises from repeated evaluations of the 
inflationary scalar power spectrum and the likelihood.
For the type of features we are considering, a complete run typically 
takes about $\mathcal{O}(10^{2})$ CPU hours.
To assess the robustness of the reconstruction, we perform multiple GA 
runs with different random seeds. 
While subdominant features show some variations across runs, we find that we 
are able to consistently recover the dominant functional forms in~$\epsilon_1(N)$. 


\subsection{Advantages of GA}\label{subsec:adv-GA}

There are several well-established methods for either adding or reconstructing 
features in the primordial scalar power spectrum.
Each method takes a different approach to balance simplicity and flexibility.
Importantly, it should be possible to physically interpret the spectrum that 
is eventually arrived at.
In this subsection, we shall highlight the advantages of GA over the other
methods.

One of the most common methods for arriving at features in the primordial scalar
power spectrum that improve the fit to the data is based on specific 
templates for the features or inflationary models.  
The method often focuses on a particular form for the feature in the primordial
spectrum, such as a sharp drop in power on large scales~\cite{Contaldi:2003zv,
Jain:2009pm,Ragavendra:2020old}, a burst of oscillations over intermediate 
scales~\cite{Hazra:2010ve} or continued oscillations over a wide range of 
scales~\cite{Aich:2011qv,Meerburg:2013cla,Meerburg:2013dla,Antony:2024vrx,
Peng:2025vda}.
These features are produced by inflationary models which contain either a 
point of inflection, a step or oscillations in the potential.  
In such cases, the underlying inflationary model is often well understood.  
The scalar power spectrum in such cases can be computed in an efficient manner, 
thereby facilitating an easy connection between the observations and physically
motivated models~\cite{Braglia:2020fms}. 
However, the major limitation of the approach is that the models are rigid. 
They can only lead to specific types of features, leaving little room to discover
unexpected signals.

On the other hand, the so-called principal component and the minimally 
parametric approaches reduce the rigidity that arises when one bases 
the analysis on specific inflationary models (in this regard, see, for 
instance, Refs.~\cite{Leach:2005av,Peiris:2009wp}).
These methods consider general deviations from the conventional power 
law form for the primordial scalar power spectrum. 
They expand the scalar power spectrum in terms of an orthogonal set of 
smooth basis functions and let the data decide which basis functions
are relevant. 
While the method permits a flexible, data-driven approach, it comes 
with trade-offs.
The method is limited in its ability to resolve sharp or localized
features.
Also, it is not always easy to directly map the mathematical basis used 
for reconstruction of the features in the primordial scalar power spectrum
to physical models of inflation.

Another approach to arrive at features in the primordial scalar power
spectrum that fit the data remarkably well is the method of reconstruction 
by direct inversion.
This approach adopts an even more model-independent route by solving the 
inverse problem.
It starts with the observed CMB spectra and works backwards to infer the 
primordial scalar power spectrum.
The most popular method in this regard is based on the RL 
algorithm that we had briefly discussed in the introductory
section~\cite{Shafieloo:2003gf,Shafieloo:2006hs,Shafieloo:2007tk, 
Nicholson:2009pi,Nicholson:2009zj,Ichiki:2009zz,Hunt:2013bha,
Hazra:2013eva,Hazra:2013xva,Hazra:2014jwa,Sohn:2022jsm}. 
In principle, this approach can reveal even significant departures from a 
nearly scale-invariant scalar power spectrum without assuming a specific template. 
However, in practice, the problem proves to be ill-posed and highly sensitive 
to noise.
Therefore, strong smoothness conditions must be imposed to stabilize the 
solution.
Although such a process makes the method robust, it comes at the cost of 
potentially erasing sharp or oscillatory features.
In fact, as we mentioned in the introduction, one of the goals of this
work is to arrive at reconstructed spectra based on the RL algorithm 
using GA.

Lastly, Bayesian non-parametric methods adopt a different approach 
by treating the primordial scalar power spectrum as a random function 
drawn from a probability distribution. 
Tools such as Gaussian processes or flexible spline models are then
used to reconstruct the spectrum (in this regard, see 
Refs.~\cite{Handley:2019fll,Liao:2019qoc,Li:2019kdj,Belgacem:2019zzu,
Krishak:2021fxp,Martinez-Somonte:2023ckq}). 
These approaches are powerful because they allow the data to guide the 
functional form and provide rigorous error bars. 
However, they also rely on prior assumptions, such as the choice of the
kernel or the so-called hyper-parameters.
Such dependence tend to bias the reconstruction toward smoother spectra,
thereby making it harder to capture fine oscillations or localized 
features. 

In contrast, the reconstruction based on GA proceeds within a grammar-defined
functional space, where elementary building blocks (such as exponentials, 
logarithms, trigonometric functions, polynomials and rational terms) are 
combined through evolutionary operations. 
This approach is not completely free of templates since the grammar imposes 
a basis of admissible functions.
Nevertheless, it allows the algorithm to dynamically generate rich functional
structures, such as DOGE or localized bumps, which are difficult 
to capture in conventional templates without fine-tuning. 
Thus, the GA effectively balances flexibility with the ability to 
interpret the result.
It retains enough structure to ensure physically meaningful inflationary 
backgrounds, while at the same time exploring a broad functional space 
in a data-driven manner. 
In such a sense, the GA pipeline complements existing methods by providing a 
grammar-guided, semi-analytical, yet highly flexible reconstruction strategy
capable of revealing novel inflationary scenarios beyond the reach of more 
restrictive techniques.  


\section{Test of the reconstruction procedure}\label{sec:test-run}

Before applying our pipeline to the real CMB data, we would like to first
test whether the GA-based method is able to reconstruct the first slow roll 
parameter~$\epsilon_1(N)$ leading to a known inflationary scalar power 
spectrum. 
As we mentioned in the introduction, it has been shown that a DOGE in the 
first slow roll parameter leads to a power spectrum that fits the Planck 2018 
and BICEP-Keck 2018 data (incorporated in the \texttt{Plik-TTTEEE+lowl+lowE+BK18} 
likelihood) very well~\cite{Antony:2022ert}.
To establish the reliability of our reconstruction pipeline, we shall now 
check whether the GA can faithfully recover the features contained in the
reference spectrum.


\subsection{Definition of \texorpdfstring{$\chi^2$}{chi-squared}}\label{subsec:def-chsq}

To quantify the extent of agreement between the 
GA-generated scalar power spectrum, say, $O_i$, and the reference dataset, 
say, $E_i$, we shall employ the so-called Pearson~$\chi^2$
statistic~\cite{Wall_Jenkins_2012}.
In other words, we shall minimize the quantity
\begin{equation}
\chi^2 = \sum_{i=1} \frac{(O_i - E_i)^2}{E_i},
\end{equation}
where the sum is over all the data points.
In the classical Pearson test, the denominator~$E_i$ represents the 
expected counts and approximates the variance in the Poisson-distributed 
data. 
In the present context, as we mentioned, $O_i$ and $E_i$ correspond to the 
GA-generated and expected values of the dimensionless scalar power spectrum. 
The denominator acts mainly as a normalization factor, allowing us to
evaluate relative differences between the predicted and expected values.

In the broader context of data analysis, one typically uses the following 
definition for the goodness of fit~$\chi^2$ (in this context, see, for 
instance, Ref.~\cite{Wall_Jenkins_2012}):
\begin{equation}
\chi^2 = \sum_{i,j}^{n}(x_{i}-\mu_{i})\mathrm{Cov}^{-1}_{ij}(x_{j}-\mu_{j}),
\end{equation}
where $x_i$ and $\mu_i$ denote the observed and predicted (by the model) 
values, respectively, and $n$ represents the number of data points.
Also, $\mathrm{Cov}_{ij}$ denotes the covariance matrix whose diagonal elements
depend on the uncertainties associated with each data point and the non-diagonal
elements describe the correlations between different data points.
However, since the current exercise is designed as a methodological validation 
rather than a statistical inference, the Pearson form suffices to provide a 
robust measure of the relative deviations between the reconstructed and 
reference power spectra. 
As lower $\chi^2$ values indicate a closer match, improvements across the
generations of GA reflect the increasing ability of the evolving functional 
forms to capture the desired target scalar power spectrum.


\subsection{Functions and priors}\label{subsec:func_test}

To reproduce the expected scalar power spectrum, we shall assume that the 
function $F(N)$ that describes the departures from the baseline behavior
of the first slow roll parameter [cf. Eqs.~\eqref{eq:e1-bl} 
and~\eqref{eq:m-fsrp}] can be expressed as
\begin{equation}
F(N)=\sum_{i=1}^{4} f_{i}(N)\label{eq:F-test},
\end{equation} 
where the functions $f_{i}(N)$ are given by
\begin{equation}
f_{i}(N)= \frac{A_{i} 
\cG_{I_i}[B_{i}(N-C_{i})]}{1+ \beta(N-C_{i})^2}
\label{eq:f-test}
\end{equation}
and $\cG_{I_i}(N)$ are the functions in the grammar. 
We should clarify that this form of the function $F(N)$ is motivated by 
the earlier work in this context~\cite{Antony:2022ert}. 
The parameters $(A,B,C)$ are varied randomly over the prior range
indicated in Tab.~\ref{tab:test}.
Note that the parameter $I$ has been introduced to identify the functions 
in the grammar.
We shall assume that the grammar consists of the two functions~$\cG_{0}(N)
=\sin(N)$ and~$\cG_1(N)=\cos(N)$.
Therefore, $I$ can take values of either zero or one.
Also, we shall fix the value of parameter $\beta$ to be such that 
$\log_{10}\beta=2.5$.
\begin{table}
\centering
\begin{tabular} {|c|c|}
\hline
Parameter &  Prior range\\
\hline
$A$ & $[0.07,1]$\\
\hline
$B$ & $[45, 55]$\\
\hline
$C$ &$[10,20]$\\
\hline
$I$ & $[0,len(grammar)]$\\
\hline
\end{tabular}
\caption{The range of priors for the different parameters in the function~$F(N)$
that modifies the behavior of the first slow roll parameter to reconstruct the 
reference scalar power spectrum [cf.  Eq~.~\eqref{eq:e1-bl} and~\eqref{eq:m-fsrp}, 
and Eqs.~\eqref{eq:F-test} and~\eqref{eq:f-test}].
Note that the parameter $I$ is an integer and can take values of either zero or 
one.}
\label{tab:test}
\end{table}

\subsection{Results of the run}\label{subsec:results-test}

\begin{figure*}
\includegraphics[width=0.475\linewidth]{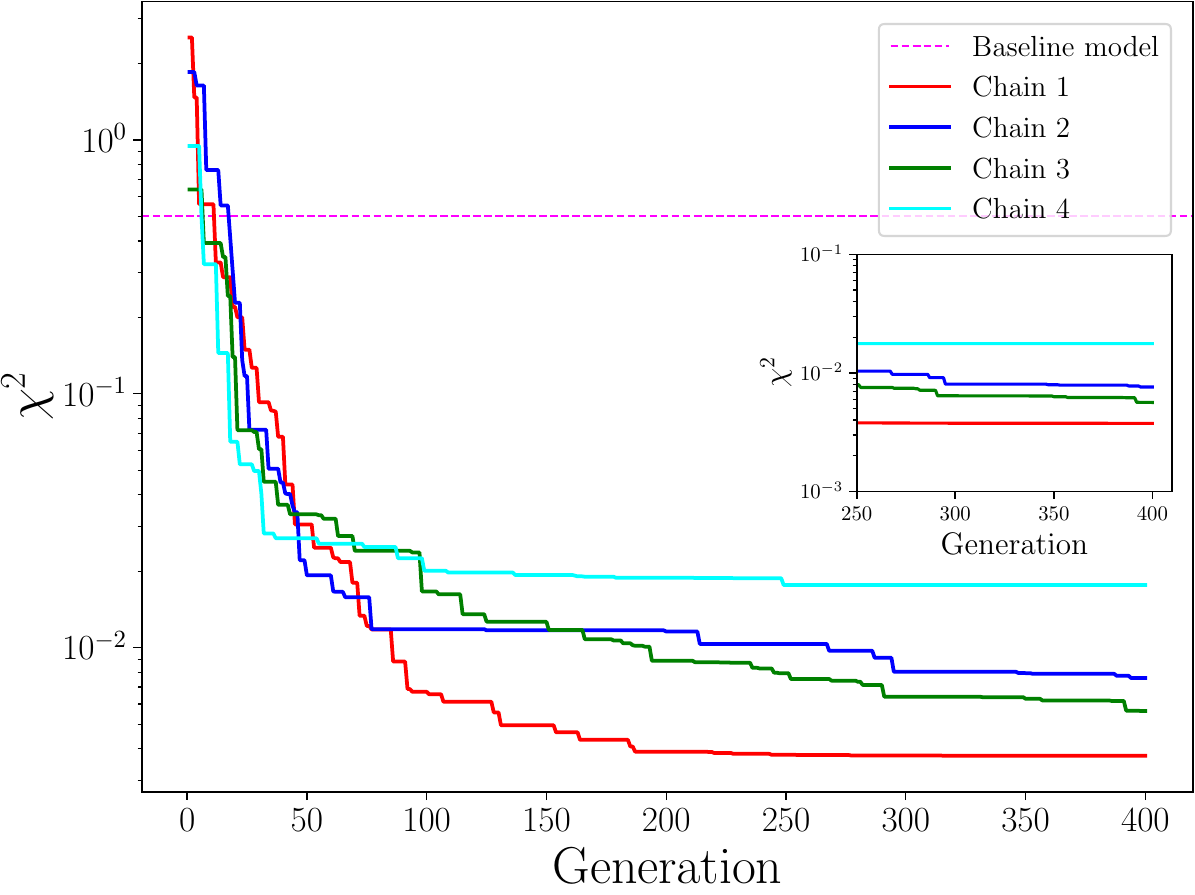}
\hskip 10pt
\includegraphics[width=0.475\linewidth]{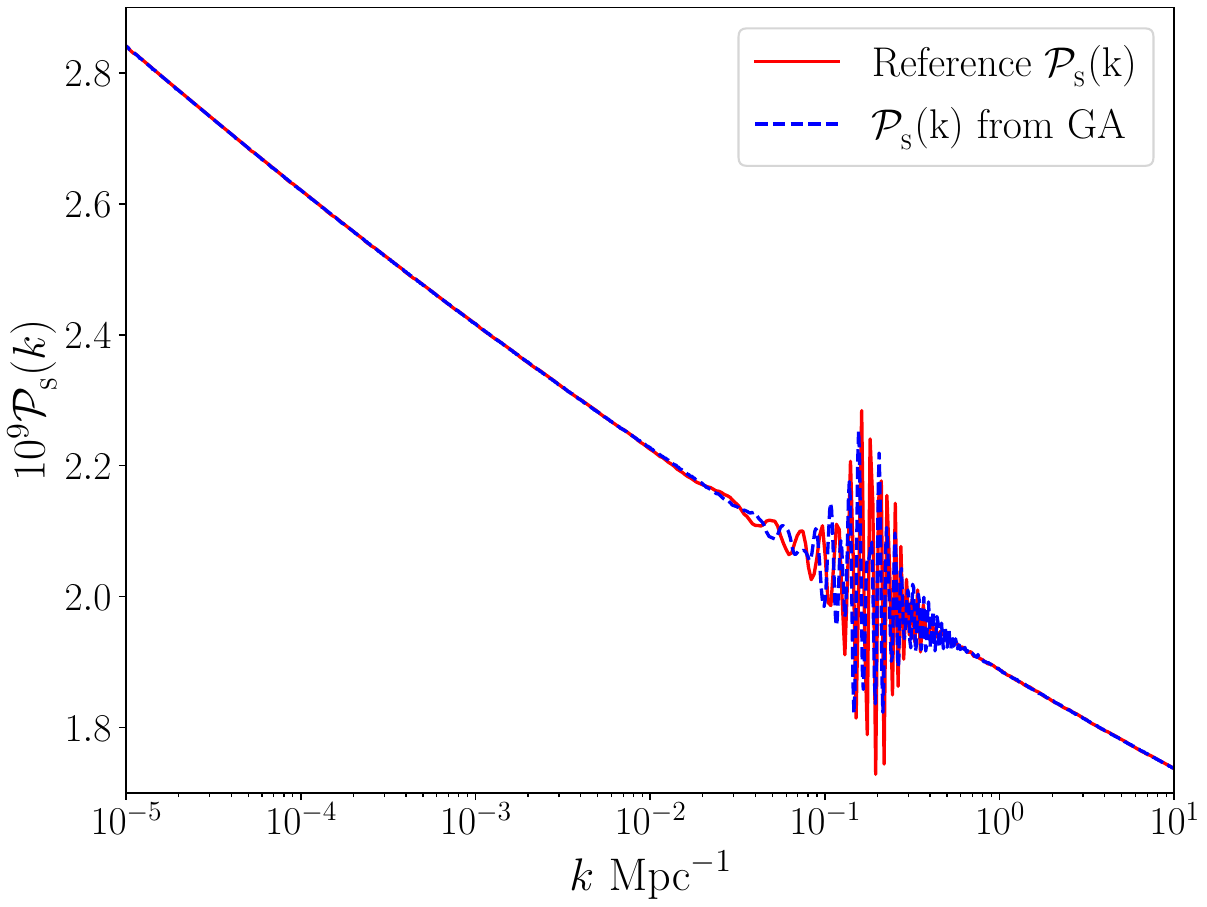}
\caption{We have illustrated the results of the exercise we have carried out
to reconstruct a given reference scalar power spectrum~$\ps(k)$ containing 
features using GA.  
We have plotted (on the left) the evolution of the fitness score or the 
goodness of fit, i.e.~$\chi^2$, across the population as a function of 
generations for four different chains.
We have also included an inset highlighting the behavior of the fitness score 
over the final $150$ generations (on the left).
Moreover, we have indicated the goodness of fit for the baseline model 
(on the left, in magenta).
Further, to illustrate the performance of the GA, we have plotted (on the right) 
the reconstructed as well as the reference inflationary power 
spectra~\cite{Antony:2022ert}.
It is clear from the plots of the power spectra that the GA-based approach is 
able to reproduce the features in the reference scalar power spectrum fairly 
well.}\label{fig:test}
\end{figure*}

We first compute the scalar power spectrum~$\ps(k)$ generated by the 
baseline first slow roll parameter given by Eq.~\eqref{eq:e1-bl}. 
The resulting scalar power spectrum, when compared with the reference 
scalar power spectrum containing features, yields the goodness of fit
value of $\chi^2 = 0.85$ over $500$~points.
This already indicates reasonable agreement with the reference spectrum, 
but leaves room for improvement. 
We then consider the power spectrum generated by the modified first slow 
roll parameter described by Eq.~\eqref{eq:m-fsrp}, with $F(N)$ given by 
Eqs.~\eqref{eq:F-test} and~\eqref{eq:f-test}.
We run the GA with an initial population of $100$ individuals, evolved 
across $400$ generations using the crossover and mutation operators, as
described in the previous section.
The algorithm identifies the best-fit modification to $\epsilon_1(N)$, 
leading to an improved value of $\chi^2 = 0.032$.
The explicit analytical form of the modification function $F(N)$ recovered 
from the GA run is given by
\begin{align}
F(N) &= 0.037
\frac{\sin[46.74 (N - 18.09)]}{[1+316.23(N - 18.09)^2]}\nn\\
&\quad+0.052 \frac{\sin[52.25 
(N - 18.04)]}{[1+316.23 (N - 18.04)^2]}\nn \\
&\quad+0.037 \frac{\cos [45.24 (N- 18.04)]}{[1
+316.23(N - 18.04)^2]}.
\label{eq:GA-test}
\end{align}
In Fig.~\ref{fig:test}, we have illustrated the performance of the GA across 
generations.
Specifically, we have plotted the fitness score or the goodness of fit
value, i.e. $\chi^2$, as it evolves with the new generations.
We observe a steep drop in $\chi^2$ during the first nearly $50$ or so 
generations, which reflects the GA’s rapid exploration of functional space.
This is followed by a slower, asymptotic improvement, with the population
stabilizing near the best-fit solution after about~$200$ generations. 
In the figure, we have also plotted the reconstructed as well as the 
reference scalar power spectra~\cite{Antony:2022ert}.
It should be clear from the figure that the reconstructed spectrum shows 
excellent agreement across all scales, including the oscillatory features 
that reflect the non-trivial dynamics of the inflationary background.

This reconstruction exercise serves as a stringent consistency check for 
the GA framework. 
The fact that the algorithm not only converges rapidly but also reproduces 
the detailed oscillatory pattern of the reference scalar power spectrum 
demonstrates both the accuracy and the robustness of our method. 
It also highlights a key advantage of the grammar-guided GA approach.
The approach has the ability to recover non-trivial functional forms of 
$\epsilon_1(N)$ without presupposing their structure, while maintaining 
theoretical consistency with inflationary dynamics. 
The successful outcome of this test provides strong justification for 
applying the pipeline to the real CMB data.


\section{Functions and priors for reconstructing spectra 
with features}\label{sec:choice-of-function}

In Sec.~\ref{sec:methodology}, we described the architecture and workflow 
of the GA pipeline and, in the previous section, we illustrated the 
effectiveness of the GA by reconstructing a previously known spectrum.
In this section, we shall discuss in some detail the three types of 
features that we aim to reconstruct using GA. 
As we discussed in the introductory section, the three types are as 
follows:~(i)~features generated by DOGE in the first slow roll 
parameter, (ii)~features generated by the CPSC model, and (iii)~features 
reconstructed using the MRL alogorthim. 

We shall begin by outlining the theoretical motivations for considering 
each type of feature and briefly discuss the manner in which they fit 
the outliers in the CMB data.
Thereafter, we shall describe the templates for the grammar that we aim 
to employ in GA for these reconstructions.
We shall also indicate the corresponding ranges in priors to be considered 
for the parameters involved. 
By combining the templates and priors with the adaptive GA framework,
our approach efficiently explores the functional space of~$\epsilon_1(N)$
and identifies the features in the inflationary scalar power spectrum that 
are  most consistent with the observations by Planck.
Such a unified and consistent setup ensures that any variations observed 
in the resulting scalar power spectra can be directly attributed to the 
functional forms of~$\epsilon_1(N)$.
Finally, we should add that, with the form of $\epsilon_1(N)$ in hand, we 
can, in principle, construct the inflationary potential that leads to the 
predicted scalar power spectrum. 


\subsection{Features generated by DOGE in 
\texorpdfstring{$\epsilon_1(N)$}{epsilon1}}\label{subsec:doge-I}

It has been repeatedly found that a localized burst of oscillations in the 
primordial scalar power spectrum can improve the fit to the Planck data (in 
this regard, see, for example, Refs.~\cite{Hazra:2010ve,Planck:2018jri}).
In a recent work, as we discussed, such oscillations were introduced in the 
primordial scalar power spectrum by assuming that the first slow roll parameter
contains DOGE~\cite{Antony:2022ert}.
Interestingly, in addition to enhancing the fit to the CMB temperature and 
polarization data, the localized features generated in such a scenario have 
the notable effect of simultaneously increasing the value of the Hubble 
constant~($H_{0}$) and decreasing the amplitude of fluctuations in the 
matter density today~($S_{8}$). 

We shall now compare the DOGE scenario with the Planck data.
To do so, we shall adopt the same template for grammar that we introduced in 
Eqs.~\eqref{eq:F-test} and~\eqref{eq:f-test}. 
We shall also retain the range of prior ranges listed in Tab.~\ref{tab:test}.
We shall assume that the feature can appear over the entire range of wave numbers 
that the CMB is sensitive to.


\subsection{Features generated by the CPSC model}\label{subsec:cpsc}

The CMB data contain two sets of prominent outliers when compared with the 
angular power spectrum corresponding to a nearly scale-invariant primordial 
scalar power spectrum (in this regard, see Fig.~\ref{fig:cmb_spectra_TE_EE}).
It has been known for a while that the CMB data point to a suppression in 
the power at the lower multipoles of $\ell < 30$ (in this regard, see,
for example, Refs.~\cite{Bridle:2003sa,Contaldi:2003zv,Cline:2003ve,
Nicholson:2007by,Hergt:2018ksk,Cicoli:2014bja,Ragavendra:2020old}).
More recently, it has been noticed that the data also point to possible
departures from the predictions of a nearly scale-invariant scalar power 
spectrum around $\ell \simeq 750$~\cite{Chen:2014cwa,Braglia:2021rej}. 
Despite numerous efforts, no single-field inflationary model has been able 
to simultaneously address these two prominent outliers observed in the CMB 
angular power spectrum.

One of the most compelling scenarios that can potentially explain both these 
outliers is the so-called CPSC model that we mentioned in the introductory 
section (in this regard, see Refs.~\cite{Chen:2014cwa,Braglia:2021rej}). 
These models lead to a sharp feature in the scalar power spectrum on large 
scales.
They also generate oscillatory features on smaller scales.
These two distinct features are generated through mechanisms that follow
one another during the early stages of inflation.
Therefore, their locations in the power spectrum and the nature of these 
features prove to be related.
A complete CPSC inflationary model involves two fields~\cite{Braglia:2021rej}.
The potential describing the two fields contains a valley which turns its
direction in the field space.
Also, the slope of the valley is steeper along one direction of the field
than the other.
Moreover, there is a step in the potential along the steeper direction.
When the fields are displaced from the minima of the potential, initially,  
the field characterized by the steeper slope rolls down the potential, 
whereas the other field hardly evolves.
As the first field crosses the step, it gains speed, and as it approaches 
the turn in the valley, the second field begins to evolve.
However, due to its relatively larger speed compared to the second field, 
after the turn, the first field slightly overshoots the minimum of the 
potential, which makes it oscillate about the minimum of the valley.
While the step in the potential and the turn in the field space generate a 
sharp feature in the scalar power spectrum, the subsequent oscillations of 
the field about the minimum of the potential give rise to an oscillatory 
feature (which is referred to as the clock signal) in the scalar power spectrum.
Further, the resonance between the oscillating background and the oscillations 
of the perturbations in the sub-Hubble domain lead to modulations in the scalar
power spectrum over a wide range of scales (in this regard, see, for instance,
Refs.~\cite{Chen:2008wn,Noumi:2013cfa,Chowdhury:2016yrh}).
We should mention that arriving at such a power spectrum in a single-field 
inflationary model has remained a challenge.

Let us now employ the GA pipeline to construct a single-field inflationary 
model that mimics the power spectrum generated by the CPSC model.
In this case, the modification function $F(N)$ that we shall work with is 
as follows:
\begin{equation}
F(N)=\sum_{i=1}^{2} f_{i}(N),\label{eq:F-cpsc}
\end{equation} 
where 
\begin{align}
f_{i} (N)
&= A_{i} \frac{{\tanh}(N-B_{i})}{1+0.99 (N-B_{i})^2}\nn \\ 
&\quad+ E_{i}\e^{-(N-D_{i})^2} \cG_{I_{i}}[F_{i}(N-D_{i})]\label{eq:f-cpsc}
\end{align}
and $\cG_{I_{i}}(N)$ are functions in the grammar. 
The choice of grammar is similar to what we discussed in Sec.~\ref{subsec:func_test}, 
i.e. we shall work with $\cG_{0}(N)=\sin(N)$ and~$\cG_1(N)=\cos(N)$.
The motivation for employing a combination of hyperbolic and trigonometric
functions in $F(N)$ is to construct a form that naturally generates the 
localized bump as well as oscillatory features in the scalar power 
spectrum.
For each population, we generate six sets of random numbers for~$(A, B, C, 
D, E)$ and integer~$I$ from the prior range listed in Tab.~\ref{tab:cpsc}.
\begin{table}
\centering
\begin{tabular} {|c|c|}
\hline
Parameter &  Prior range\\
\hline
$A$ & $[-0.028,-0.009]$\\
\hline
$B$ & $[16,18]$ \\
\hline
$C$ & $[17,19]$\\
\hline
$D$ & $[0.001,0.005]$\\
\hline
$E$ & $[10,20] $\\
\hline
$I$ & [0,len(grammar)]\\
\hline
\end{tabular}
\caption{The range of priors for the different parameters that describe the
function~$F(N)$ which modifies the behavior of the baseline first slow roll 
parameter in the CPSC model [cf. Eq.~\eqref{eq:m-fsrp}, and 
Eqs.~\eqref{eq:F-cpsc} and~\eqref{eq:f-cpsc}].}
\label{tab:cpsc}
\end{table}

We shall see that, from such a template, our GA pipeline is able to arrive 
at features very similar to those generated by the CPSC model.
As we mentioned, the CPSC model involves two fields.
In contrast, using the GA pipeline we are able to generate similar features 
in what is effectively a suitably tuned single-field inflationary model.


\subsection{Features reconstructed by MRL}\label{subsec:mrl}

We shall now discuss the third type of feature of interest, viz. the 
primordial power spectrum reconstructed from the CMB data.
The so-called RL algorithm was originally developed for reconstructing 
images in astrophysics~\cite{Richardson:72}.
Over the years, it has found extensive applications in cosmology. 
As we discussed in the introductory section, the MRL algorithm has been employed 
to reconstruct the primordial scalar power spectrum from the observed CMB 
angular power spectra~\cite{Shafieloo:2003gf,Shafieloo:2006hs,Shafieloo:2007tk,
Nicholson:2009pi,Nicholson:2009zj,Hazra:2013eva,Hazra:2013xva,Hazra:2014jwa}.
Let $\Delta_\ell^{X}(k)$ denote the transfer functions that relate the
primordial scalar power spectrum $\ps(k)$ to the angular power spectra of
the CMB, say, $C_{\ell}^{XX}$, where $X$ represents either the 
temperature $T$ or the $E$-mode polarization of the CMB [cf. Eq.~\eqref{eq:Cl}].
The transfer functions~$\Delta_\ell^{X}(k)$ depend on the background 
cosmological parameters.
Given the transfer function, a {\it regularized}\/ MRL algorithm has been 
used recently to reconstruct the primordial scalar power spectrum from the 
Planck CMB data~\cite{Sohn:2022jsm}.
We should clarify that the term regularized refers to a data analysis
technique that reduces the amplification of noise in the reconstructed 
power spectrum.
However, to our knowledge, no inflationary model (either single or multi-field) 
has been proposed to generate such a power spectrum.
Inspired by the reconstruction, we aim to implement the GA to arrive at
an effective single-field inflationary model that reproduces a similar 
feature-rich scalar power spectrum.

To mimic the amplitude of the suppression seen in the MRL-reconstructed 
scalar power spectrum on large scales, we find it necessary to slightly 
modify the parameters of the baseline, slow roll, inflationary model.
Instead of the values we mentioned earlier, we assume the values of the 
parameters describing the baseline first slow roll parameter to be 
$(\epsilon_{1a}, \epsilon_{2a}, N_1)=(9.64\times 10^{-3}, 3.23\times 
10^{-2}, 20)$ [see Eq.~\eqref{eq:e1-bl}].
To reproduce the desired features, we consider $F(N)$ to be composed of 
only two terms as in the CPSC model.
Nevertheless, each term still contains sufficient freedom to capture the 
key features as well as reflect the complexity of the reconstructed model.
The overall $F(N)$ to be reconstructed by the GA is assumed to be of the
form
\begin{equation}
F(N)=-1+\sum_{i=1}^{2} f_{i}(N),\label{eq:F-mrl}
\end{equation} 
where the functions $f_i(N)$ are given by
\begin{align}
f_{i}(N)&=1-\mathrm{tanh}\l(\f{N-A_{i}}{B_{i}}\r)\nn\\
&\quad+0.042 \l[1 + \f{C_{i}}{1+D_{i}(N-E_{i})^2}\r]\nn\\ 
&\quad+0.048 \l\{1 + 0.62 
\cG_{I_{i}}\l[\f{\mathrm{G}_{i}(N -F_{i})}{1+31 (N-F_{i})^2}\r]\r\},\label{eq:f-mrl}
\end{align}
with $\cG_{I_i}(N)$ representing the functions in the grammar. 
We adopt the same grammatical choice as in Sec.~\ref{subsec:func_test}, where,
recall that, we had set $\cG_0(N)=\sin(N)$ and $\cG_1(N)=\cos(N)$.
For each population, we generate seven sets of random numbers for~$(A, B, C, 
D, E, F, G )$ and $I$ each from the predefined prior range listed in Tab.~\ref{tab:mrl}.
\begin{table}
\centering
\begin{tabular}{|c|c|}
\hline
Parameter &  Prior range\\
\hline
$A$ & $[13.5,15]$\\
\hline
$B$ & $[0.15, 0.4]$\\
\hline
$C$ & $[0.95,1.1]$\\
\hline
$D$ & $ [200,700]$\\
\hline
$E$ & $[14,16]$\\
\hline
$F$ & $[15,20]$\\
\hline
$G$ & $ [200,700]$\\
\hline
$I$ &  [0,len(grammar)]\\
\hline
\end{tabular}
\caption{The range of priors for the different parameters that describe 
the function $F(N)$ that modifies the behavior of the baseline first 
slow roll parameter in the MRL case [cf. Eq.~\eqref{eq:m-fsrp}, and 
Eqs.~\eqref{eq:F-mrl} and~\eqref{eq:f-mrl}].}\label{tab:mrl}
\end{table}


\section{Primordial and CMB spectra}\label{sec:results}

Before presenting our results, let us clarify a couple of points concerning the 
parameters and priors that we have worked with.
In the first part of the results that we shall present below, we have fixed the 
background cosmological parameters to be that of the standard $\Lambda$CDM best-fit 
values, which are arrived at assuming a nearly scale-invariant primordial scalar 
power spectrum. 
These parameter values are as follows: 
$(\Omega_{\mathrm{b}} h^2, \Omega_{\mathrm{c}} h^2, \theta, \tau)
=(0.0223,0.1201,1.041,0.05430)$~\cite{Planck:2018vyg}. 
In addition, we should mention that the derived values for $H_0$ and the $S_8$ 
corresponding to the best-fit $\Lambda$CDM model and a nearly scale-invariant 
primordial scalar power spectrum are $H_0 = 67.32\, \mathrm{km}\,\mathrm{s}^{-1}\,
\mathrm{Mpc}^{-1}$ and $S_8 = 0.812$.
As we indicated, we have run our analysis for an initial population of 
$100$~$\epsilon_1(N)$.
These populations were evolved over $400$ generations. 
We have worked with the set of GA parameters, viz. the choice of priors, 
crossover and mutation probabilities and selection criteria, that we  
mentioned in the previous three sections.


\subsection{Best-fit inflationary scalar power spectra}\label{subsec:results}

In this section, we shall present the main results of our analysis.
We shall illustrate the manner in which the fitness score or the goodness of
fit~$\chi^2$ improves across generations.
We shall also illustrate the behavior of the background, specifically the first
slow roll parameter $\epsilon_1(N)$, and the resulting scalar power spectrum for
the best-fit values in the three cases of interest.


\subsubsection{In the DOGE scenario}\label{subsec:res-doge-I}

\begin{figure*}
\includegraphics[width=0.475\textwidth]{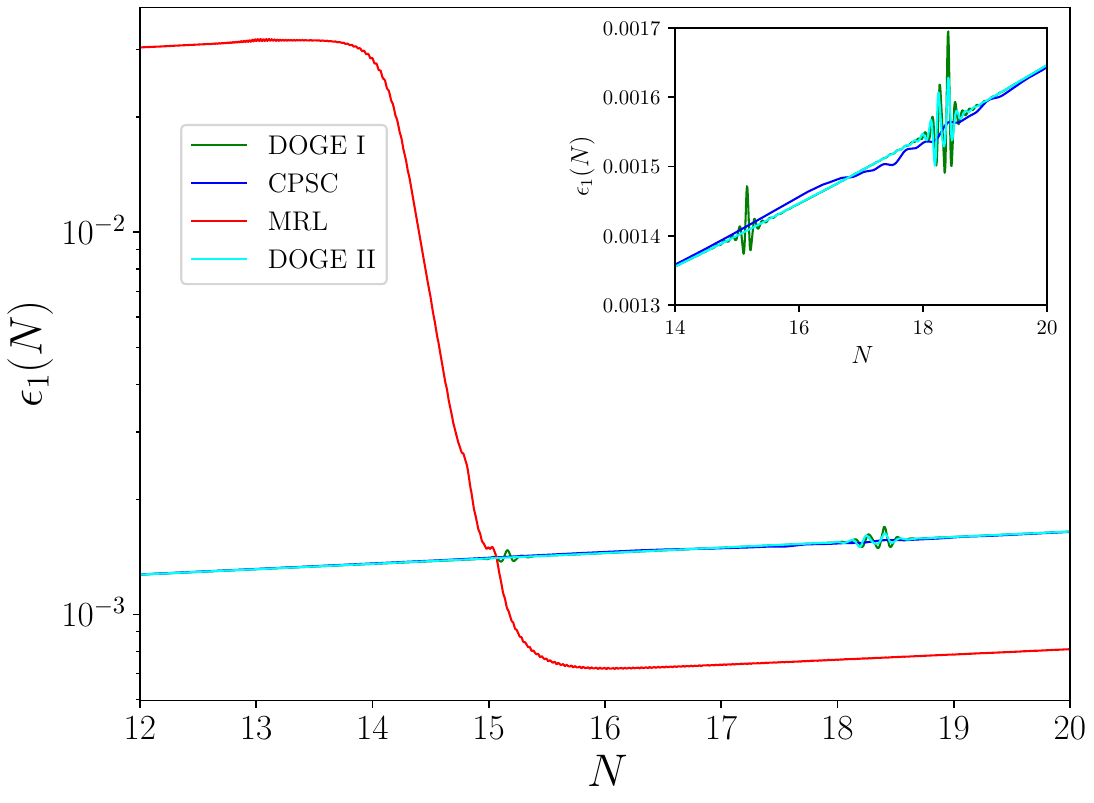}
\hskip 10pt
\includegraphics[width=0.475\textwidth]{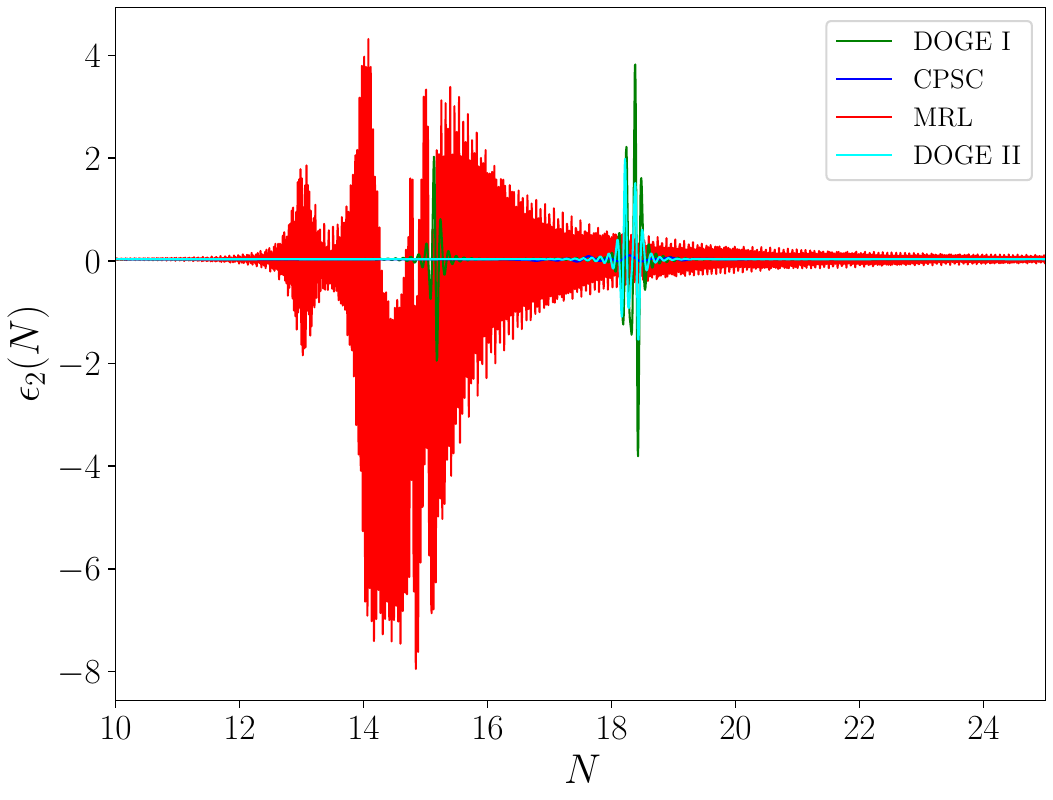}
\caption{The behavior of the first and the second slow roll parameters, viz.
$\epsilon_1(N)$ and $\epsilon_2(N)$, reconstructed by the GA have been plotted 
(on the left and right, respectively) in all the cases of interest.
To highlight the features identified by the GA, we have plotted the slow roll
parameters within the range of $e$-folds $12 < N < 20$. 
We have also included an inset (in the figure on the left) to highlight the features
in the $\epsilon_1(N)$ in the scenarios of DOGE~I and DOGE II (in this regard, see
the following section)  as well as the CPSC model.}
\label{fig:eps_1_eps_2}
\end{figure*}

\begin{figure*}
\includegraphics[width=0.475\textwidth]{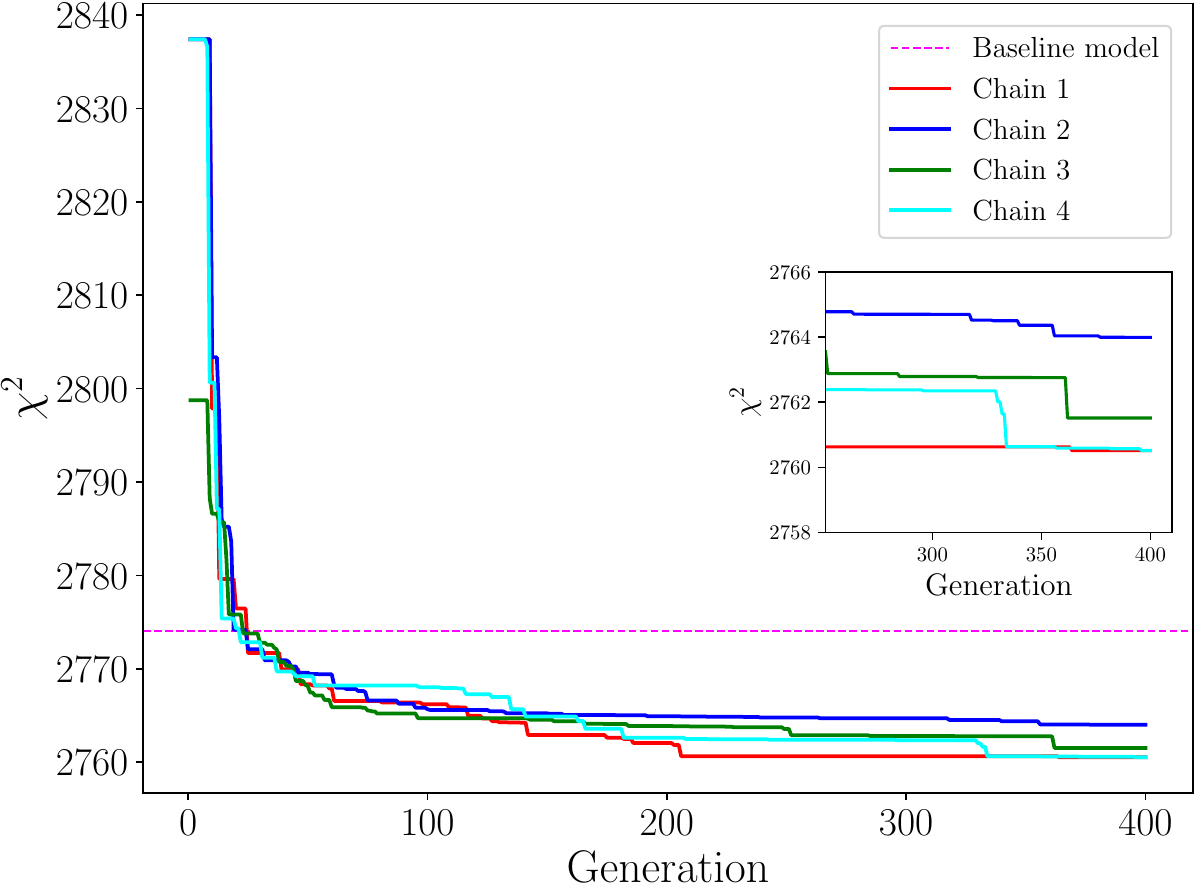}
\hskip 10pt
\includegraphics[width=0.475\textwidth]{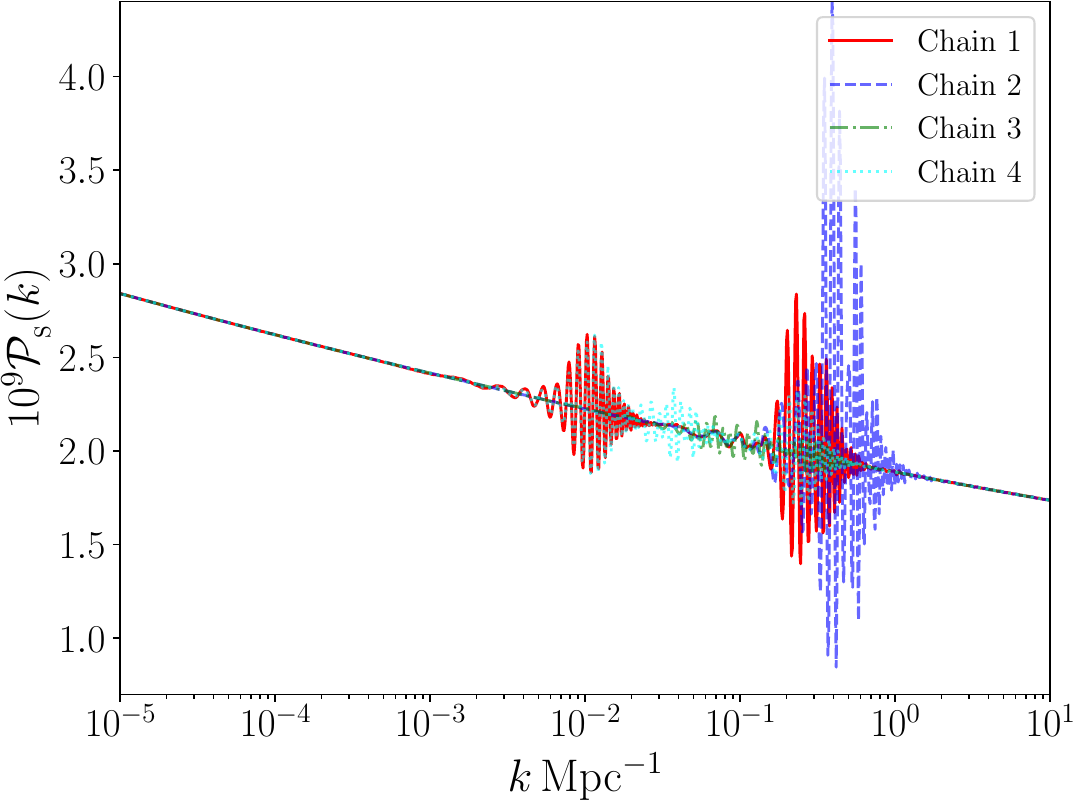}
\caption{We have plotted the evolution of the fitness score $\chi^2$ as a function 
of generations in the DOGE~I scenario for four different chains (on the left, in 
red, blue, green, cyan).
We have also plotted the corresponding scalar power spectra that lead to the best-fit 
to the Planck 2018 data for each of these chains (on the right, in the same colors).
Moreover, we have included an inset highlighting the behavior of the fitness score 
over the final $150$ generations (on the left).
Further, we have indicated the baseline fitness score of $\chi^2=2774.09$ (on the 
left, in magenta). 
The power spectrum containing two bursts of oscillations with a roughly Gaussian 
envelope (on the right, in red) leads to the best-fit among the four chains.}
\label{fig:doge-I}
\end{figure*}

Let us first consider the case wherein the baseline first slow roll parameter
is modified by DOGE.
(Later, we shall consider another set of background parameters for DOGE.
Therefore, for convenience, we shall refer to this case as DOGE~I.)
In this case, we obtain the minimum value of the goodness of fit to be~$\chi^2 
= 2760.51$.
This corresponds to an improvement of $\Delta\chi^2=-13.57$ when compared to the 
standard nearly scale-invariant scalar power spectrum.
Also, we find that, the modification function~$F(N)$ that leads to the best-fit
to the Planck data is given by
\begin{align}
F(N) &= 0.046 \f{\cos[51.00 (N - 15.16)]}{[1+316.23 (N - 15.16)^2]}\nn\\
&\quad+0.029 \f{\sin[54.93 (N - 18.25)]}{[1+316.23 (N - 18.25 )^2]}\nn\\
&\quad+0.029 \f{\cos[45.37 (N - 18.25)]}{[1+316.23 (N - 18.25 )^2]}\nn\\
&\quad+0.085 \frac{\cos[53.87 (N - 18.40)]}{[1+316.23 (N - 18.40)^2]}.
\end{align}
It should be clear from this form of $F(N)$ that the strongest modifications
arise around the $e$-folds of $N\simeq 15$ and $N \simeq 18$.
These are also evident in Fig.~\ref{fig:eps_1_eps_2} where we have plotted the 
behavior of the first and the second slow roll parameters, viz. $\epsilon_1(N)$ 
and $\epsilon_2(N)$.
The non-trivial evolution of the background affects the scalar perturbations over
scales which leave the Hubble radius during this period, imprinting specific 
features on the scalar power spectrum.
Interestingly, we find that a DOGE in the first slow roll parameter leads 
to a DOGE in the scalar power spectrum.
This aspect should be clear from Fig.~\ref{fig:doge-I} where we have
plotted the scalar power spectrum arising in the scenario.
In the figure, we have also illustrated the manner in which the fitness score
$\chi^2$ improves across generations. 


\subsubsection{In the CPSC model}\label{subsec:res-cpsc}

\begin{figure*}
\includegraphics[width=0.475\textwidth]{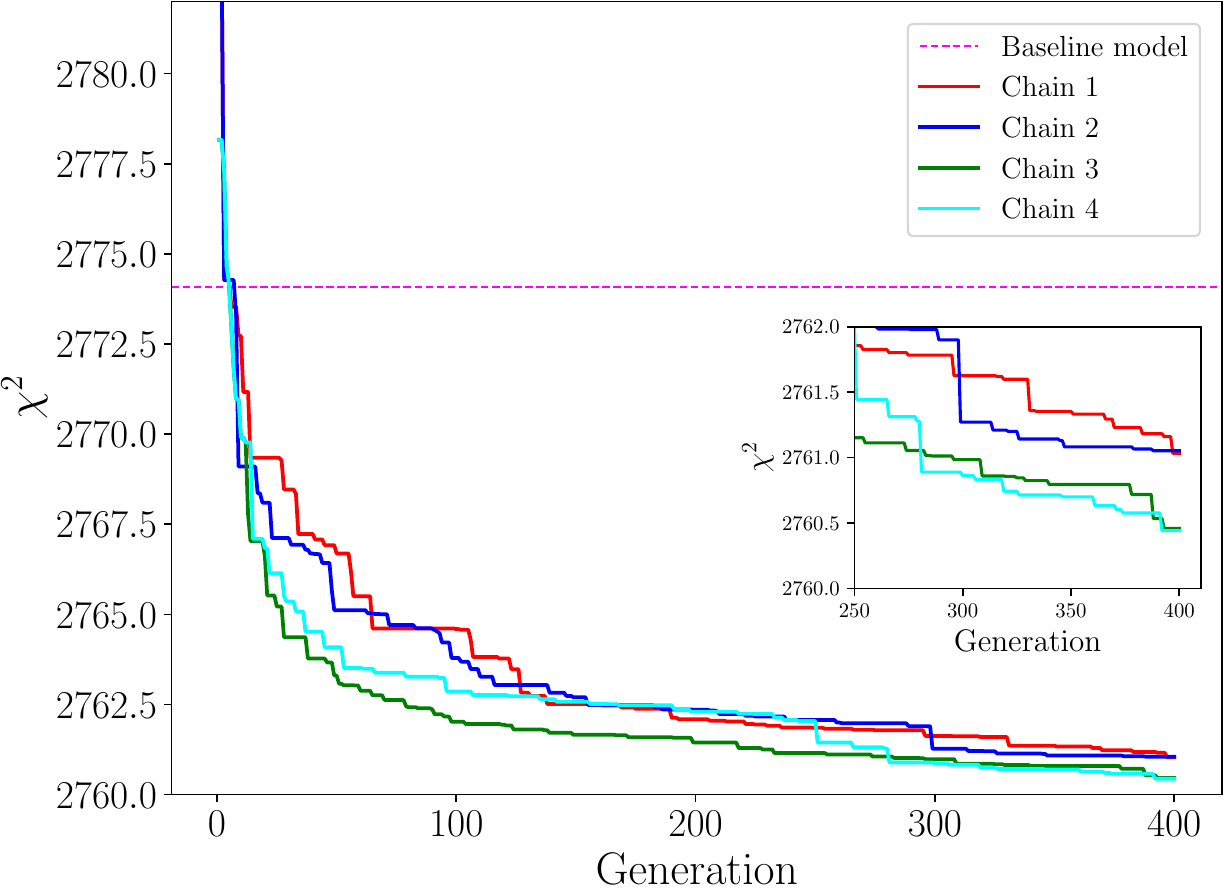}
\hskip 10pt
\includegraphics[width=0.475\textwidth]{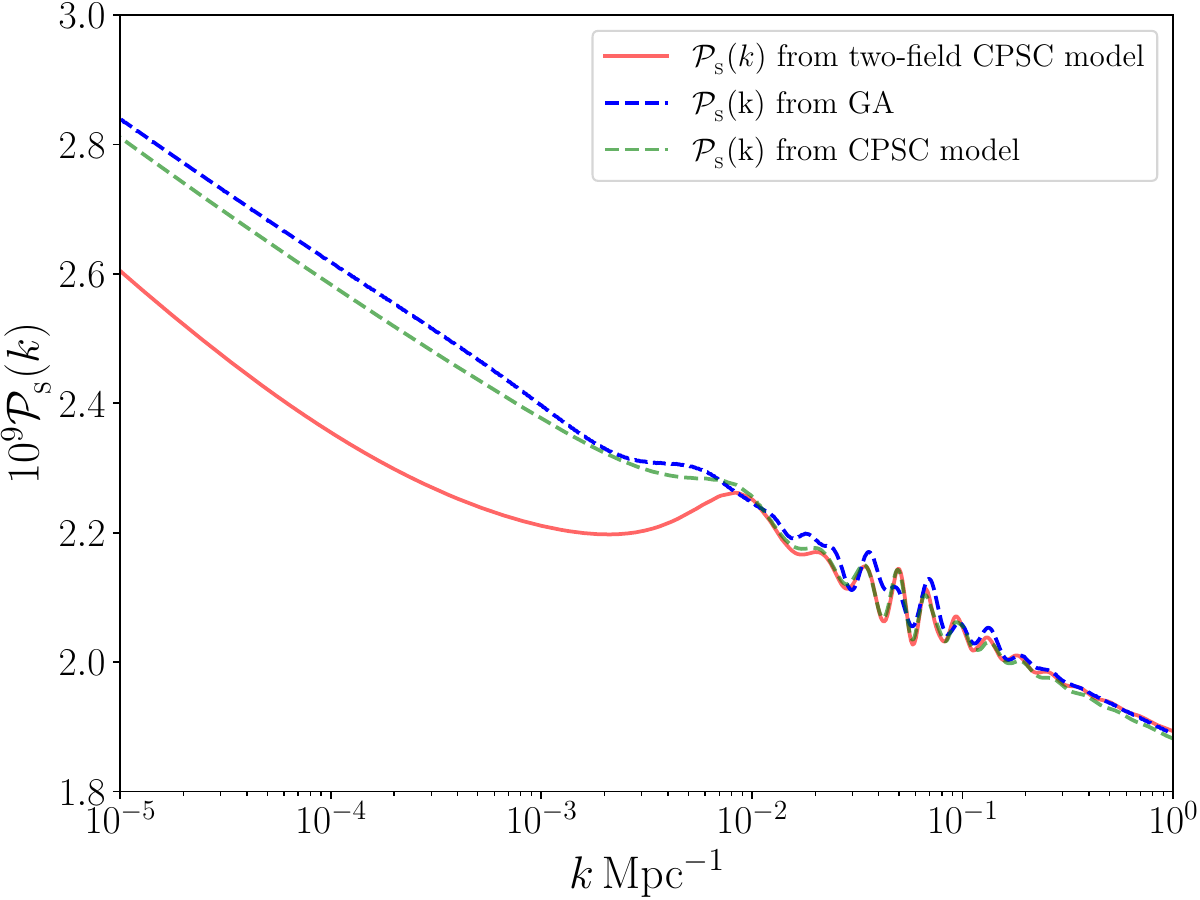}
\caption{We have illustrated the evolution of the fitness score $\chi^2$ as a function 
of generations in the CPSC model for four different chains (on the left), as in the 
previous figure.
We have included an inset highlighting the fitness score over the final $150$
generations and we have also indicated the fitness score associated with the 
baseline model (on the left), as in Figs.~\ref{fig:test} and~\ref{fig:doge-I}.
Moreover, we have plotted scalar power spectrum that leads to the best-fit to
the Planck 2018 data (on the right, in blue).
In addition, we have plotted the best-fit power spectra that have been arrived at 
earlier in specific CPSC models (on the right, in red and green)~\cite{Braglia:2021rej}.
}\label{fig:cpsc}
\end{figure*}
For the type of features generated by the CPSC models, the function~$F(N)$ that 
leads to the best-fit to the Planck 2018 data, as constructed by GA, turns out 
to be
\begin{align}
F(N)&=-0.009 \frac{{\mathrm{tanh}} [N-16.81]}{1+0.99 (N-16.81)^2}\nn\\
&\quad+0.003\, \e^{-(N-18.28)^2}{\mathrm{sin}}[10.09 (N-18.28)]\nn\\
&\quad-0.009 \frac{{\mathrm{tanh}} [N-16.84]}{1+0.99 (N-16.84)^2}\nn\\
&\quad+ 0.002\, \e^{-(N-17.96)^2}{\mathrm{sin}}[18.43 (N-17.96)].
\end{align}
In Fig.~\ref{fig:eps_1_eps_2}, we have plotted the resulting behavior of the first 
two slow roll parameters.
And, in Fig.~\ref{fig:cpsc}, we have illustrated the evolution of fitness score
$\chi^2$ across generations, when the power spectra are compared with the data.
The best-fit $\epsilon_1(N)$ leads to a goodness of fit $\chi^2= 2760.44$, which 
corresponds to an improvement of $\Delta\chi^2=-13.64$ with respect to the baseline
model.
In Fig.~\ref{fig:cpsc}, we have also plotted the scalar power spectrum associated 
with the best-fit values.
Moreover, we have compared the best-fit power spectrum we obtain with similar power
spectra arrived at earlier in the literature from the CPSC models (in this regard,
see Refs.~\cite{Chen:2014cwa,Braglia:2021rej}).
It should be clear that it is the hyperbolic tangent function in the best-fit $F(N)$ 
above that generates the small bump around $k \simeq 10^{-3}\, \mathrm{Mpc}^{-1}$ in 
the scalar power power spectrum.
Further, it should be evident that the trigonometric functions in $F(N)$ lead to 
the oscillations near $k\simeq 10^{-2}\, \mathrm{Mpc}^{-1}$. 


\subsubsection{In the MRL case}\label{subsec:res-mrl}

\begin{figure*}
\includegraphics[width=0.475\textwidth]{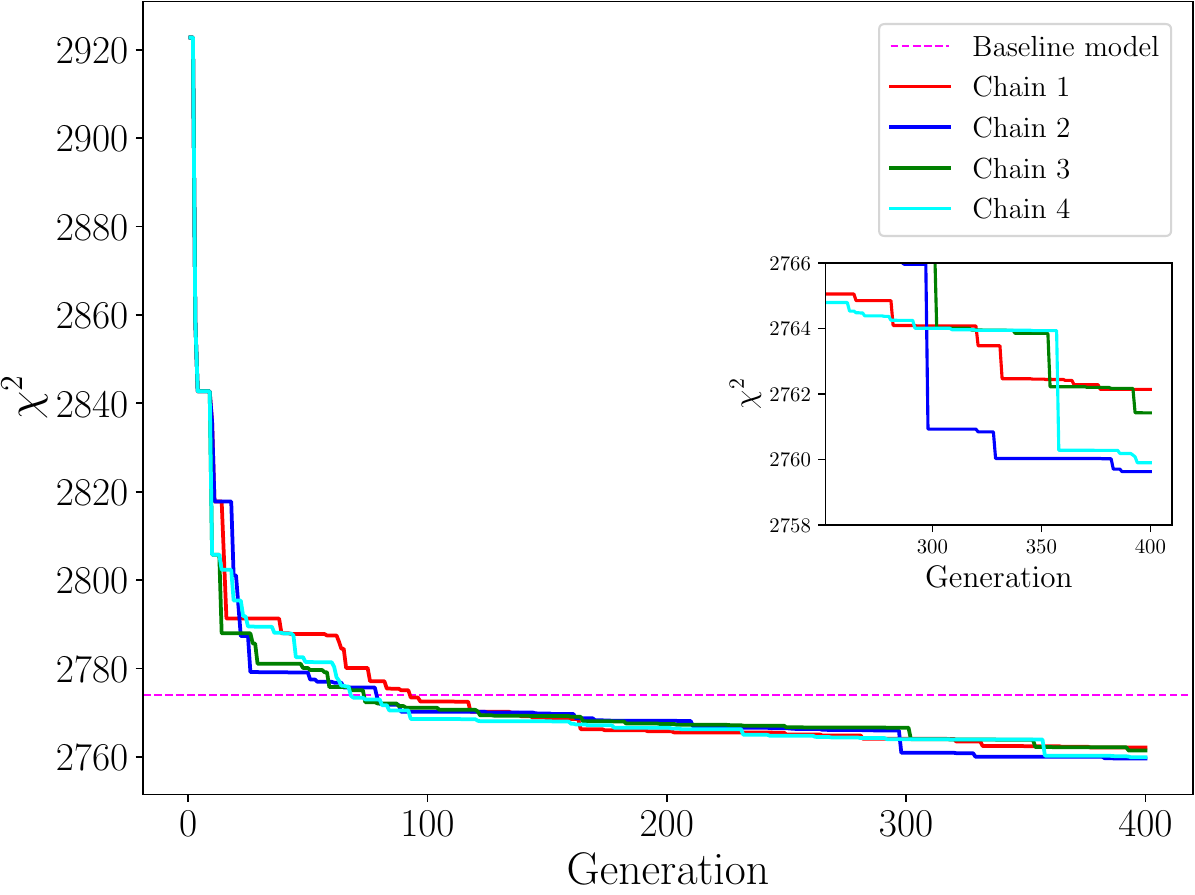}
\hskip 10pt
\includegraphics[width=0.475\textwidth]{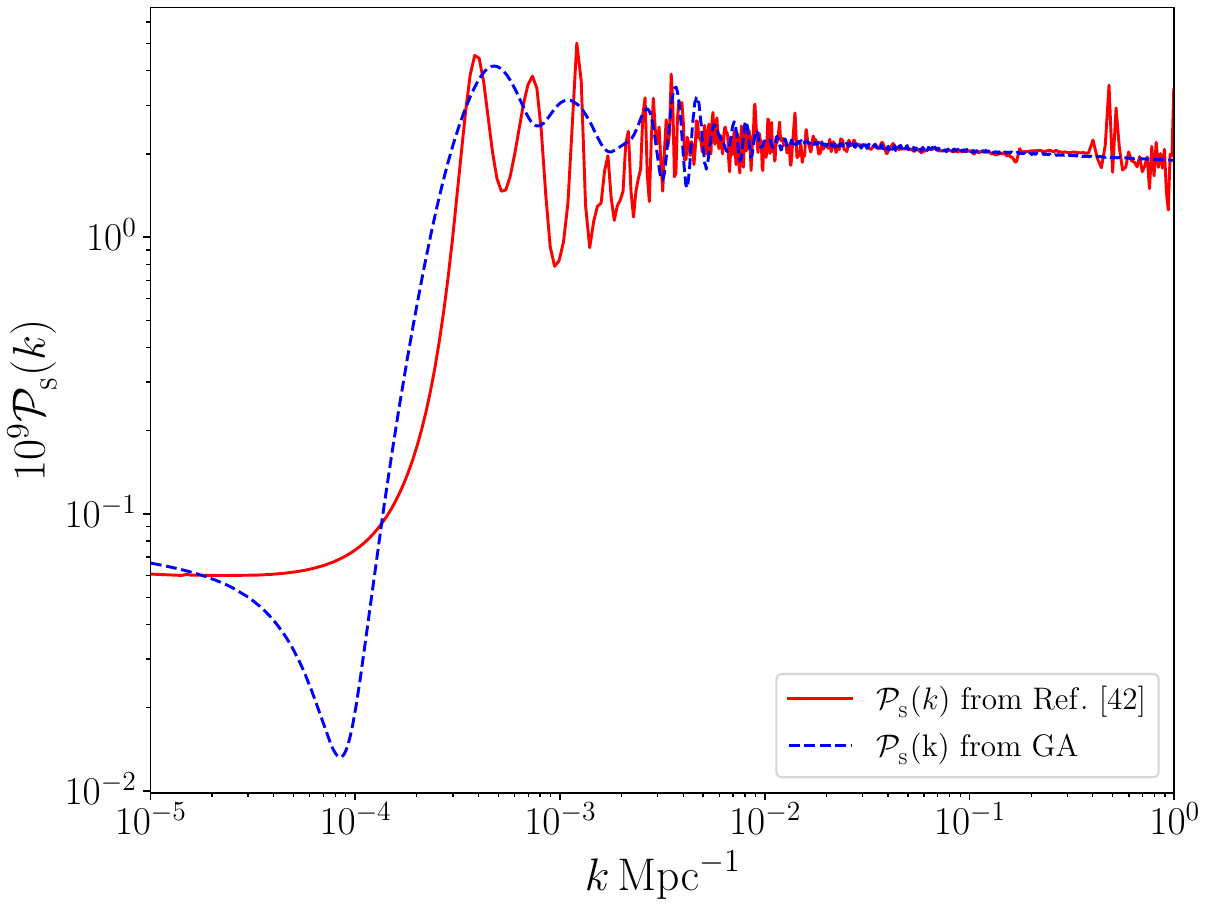}
\caption{We have illustrated the evolution of the fitness score $\chi^2$ across 
generations for four different chains to obtain features similar to the ones 
arrived at using the MRL algorithm (on the left).
As in the earlier figures, we have included an inset highlighting the fitness
scores over the final $150$ generations and we have also indicated the fitness 
score associated with the baseline model (on the left). 
Moreover, we have plotted scalar power spectrum that leads to the best-fit to
the Planck 2018 data (on the right, in blue).
Further, we have included the best-fit power spectra that have been arrived at 
earlier in a specific MRL reconstruction (on the right, in red)~\cite{Sohn:2022jsm}.}
\label{fig:mrl}
\end{figure*}

To emulate features indicated by the MRL reconstructions, we 
employ a GA run using the $F(N)$ in Eqs.~\eqref{eq:F-mrl} and~\eqref{eq:f-mrl},
along with the priors listed in Tab.~\ref{tab:mrl}.
We find the best-fit $F(N)$ to be
\begin{align}
F(N)&=1-\mathrm{tanh}\l(\f{N-14.23}{0.22}\r)\nn\\
&\quad+0.042\l[1.0 + \f{1.09}{1+342.14(N-14.80)^2}\r]\nn\\
&\quad+0.048 \l\{1.0 + 0.62\, 
\mathrm{sin}\l[\f{262.95(N-13.03)}{1+31(N-13.03)^2}\r]\r\}\nn\\
&\quad+1-\mathrm{tanh}\l(\f{N-14.32}{0.40}\r)\nn\\
&\quad+0.042 \l[1.0 + \f{0.95}{1+222.67(N-15.04)^2}\r]\nn\\
&\quad+0.048 \l\{1.0 
+ 0.62\,\mathrm{sin}\l[\f{661.21(N-14.05)}{1+31(N-14.05)^2}\r]\r\}.
\end{align}
As discussed in Sec.~\ref{subsec:mrl}, we have worked with a different 
set of parameters for the baseline model in this case.
We have plotted the corresponding behavior of the first and second roll
parameters in Fig.~\ref{fig:eps_1_eps_2}.
The choice of parameters we have worked with and the hyperbolic tangent 
function in $F(N)$ above lead to the suppression in the power spectrum
on large scales.
Also, the trigonometric functions in $F(N)$ can be expected to lead to 
oscillations in the power spectrum.
These aspects should be clear in Fig.~\ref{fig:mrl} wherein we
have illustrated the evolution of the fitness score $\chi^2$ across generations
as well as the resulting power spectrum.
In the figure, for comparison, we have included the scalar power spectrum
reconstructed using the MRL algorithm (from the data provided by the authors
of Ref.~\cite{Sohn:2022jsm}).
Although, it is rather challenging to reproduce the exact features in the power
spectrum reconstructed using the MRL algorithm, with the help of the GA, we 
are able to capture the bump and the oscillations in the power spectrum, 
resulting in a significant improvement in $\Delta\chi^2$.
This highlights GA’s ability to capture intricate and localized features in 
the inflationary dynamics. 
We should mention that the resulting best-fit modification to $\epsilon_1(N)$ 
achieves a goodness of fit of $\chi^2=2759.90$. 
This corresponds to an improvement of $\Delta\chi^2 =-14.45$ with respect to the 
baseline model.
We have carried out our analysis using the Planck likelihood module~\texttt{Plik}
in this work.
It would be worthwhile to instead calculate the improvement in the goodness of 
fit $\chi^2$ using the~\texttt{CamSpec} likelihood, which was arrived at on a 
re-analysis of the Planck 2018 data~\cite{Efstathiou:2019mdh}.


\subsection{Angular power spectra of the CMB}\label{subsec:angular-ps}

With the primordial power spectrum constructed using GA at hand, the next 
step is to examine its implications for the anisotropies in the temperature 
and polarization of the CMB.
At the level of two-point functions, the link between the primordial 
fluctuations and the anisotropies in the CMB is characterized by the angular 
power spectra of the CMB.

On fixing the background cosmological parameters, we can arrive at the 
angular power spectra of the CMB through the following relation (in this
context, see, for instance, the textbooks~\cite{Weinberg:2008zzc,
Dodelson:2020bqr,Durrer:2020fza}):
\begin{equation}
C_{\ell}^{XY} = 4\pi \int \frac{\d k}{k}\ps(k)
\Delta_{\ell}^{X}(k)\Delta_{\ell}^{Y}(k),\label{eq:Cl}
\end{equation}
where $(X, Y) = (T, E)$ denote the temperature and $E$-mode polarization, 
respectively. 
As we mentioned earlier, the quantities $\Delta_{\ell}^{X}(k)$ denote the 
transfer functions, which determine the evolution of the primordial perturbations.
We compute the angular power spectra of the CMB using the Boltzmann code
\texttt{CLASS}~\cite{Lesgourgues:2011re,Blas:2011rf}.

\begin{figure*}
\includegraphics[width=0.975\textwidth]{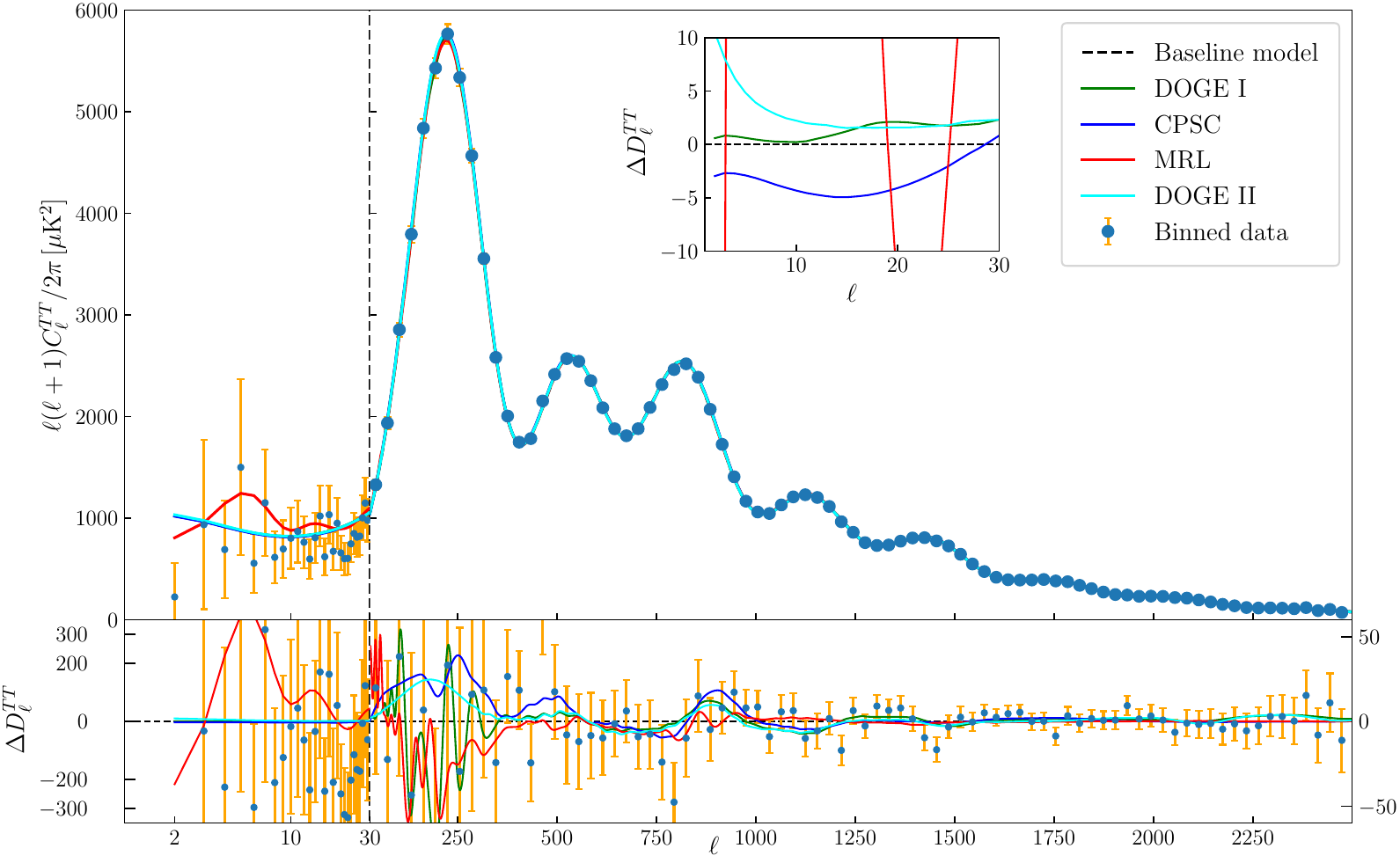}
\caption{The temperature-temperature angular power spectrum of the anisotropies 
in the CMB, i.e. $C_\ell^{TT}$, corresponding to the different GA-generated 
inflationary power spectra with features have been plotted (in the top panel, 
in red, blue, green and cyan).
In addition to the spectra from DOGE~I, CPSC and MRL, we have also illustrated 
the angular power arising in the DOGE~II scenario.
We have also included the Planck unbinned  data (over $\ell>30$) as well as
the binned data. 
Moreover, we have displayed the residuals for each primordial spectra with 
features relative to the data (in the lower panel). 
For comparison, the $TT$ spectrum derived from a nearly, scale-invariant
inflationary power spectrum generated by the baseline model is also shown 
(in the top panel).
This highlights the improvements achieved by the GA-generated features 
across different multipole ranges.
To distinguish the features of DOGE~I, CPSC, MRL, and DOGE~II, we have included
an inset (in the top panel) containing the angular power spectra over the 
lower multipoles of $\ell < 30$.}\label{fig:cmb_spectra_TT}
\end{figure*}

\begin{figure*}
\includegraphics[width=0.485\textwidth]{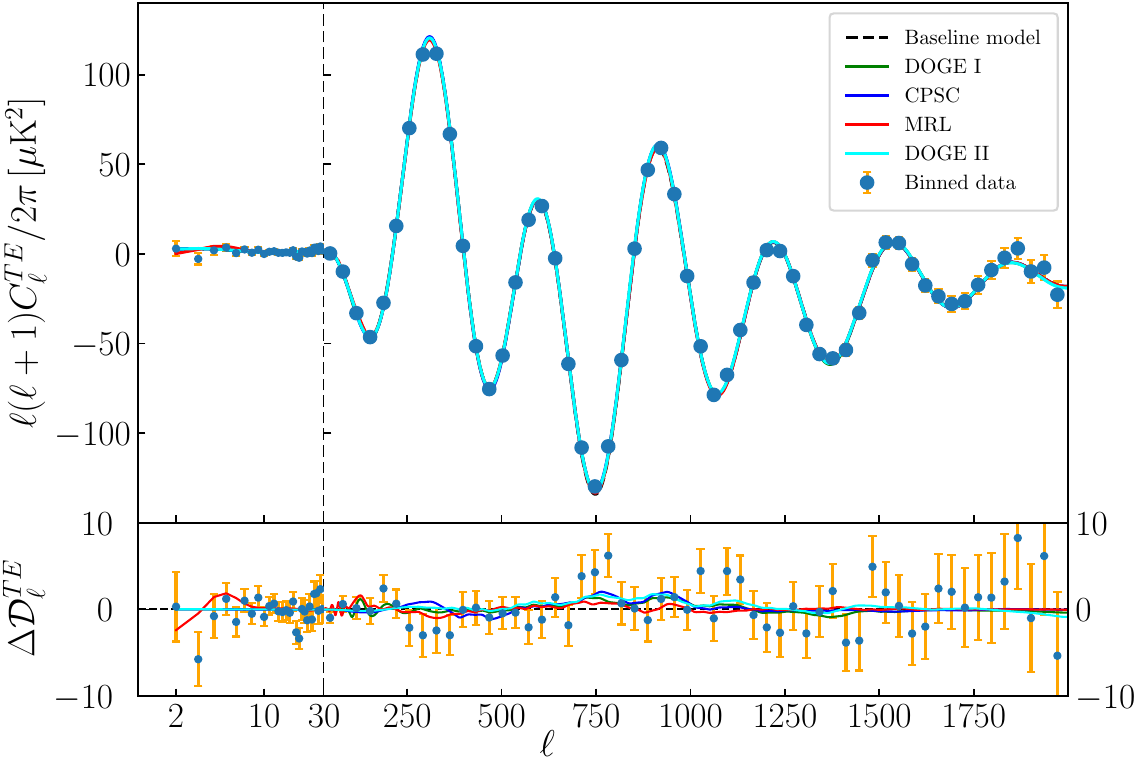}
\hskip 10pt
\includegraphics[width=0.485\textwidth]{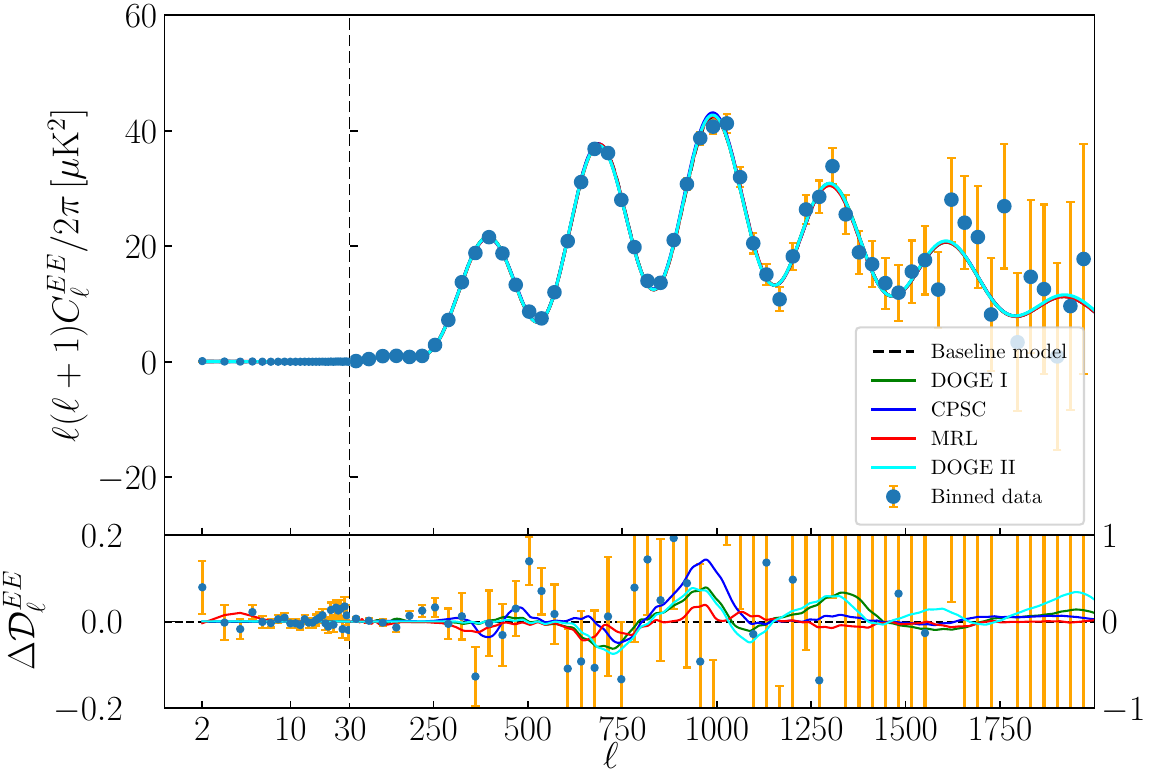}
\caption{The temperature-$E$-mode polarization and the $E$-mode polarization
angular power spectra of the anisotropies in the CMB, i.e. $C_\ell^{TE}$
and $C_\ell^{EE}$, corresponding to the different GA-generated inflationary
power spectra with features have been plotted (on the left and right, respectively,
with same choice of colors in the previous figure).
As in the previous figure, we have included the Planck data as well as the 
residuals.}\label{fig:cmb_spectra_TE_EE}
\end{figure*}

In Fig.~\ref{fig:cmb_spectra_TT}, we have illustrated the CMB 
temperature–temperature~($TT$) angular power spectra derived 
from our best-fit GA models.
These are overlaid with binned observational data from Planck. 
Similarly, in Fig.~\ref{fig:cmb_spectra_TE_EE}, we have presented the 
temperature-E-mode polarization ($TE$) and E-mode-E-mode polarization ($EE$) 
CMB angular power spectra for the same set of best-fit models, together 
with the corresponding Planck data. 
To highlight the improvements in $C_{\ell}$ introduced by the features compared 
to the standard scalar power spectrum, in the two figures, we have also included
the residuals for each GA-based model relative to the best-fit nearly, 
scale-invariant spectrum generated by the baseline model.
These residuals allows for a clear visual identification of specific multipole
ranges where the reconstructed features lead to an improved fit to the data or 
help capture known outliers in the CMB power spectra.

Let us now discuss the performance of each case in some detail.
In the DOGE~I scenario, we observe a significant improvement in the behavior of
outliers in the CMB $TT$ angular power spectrum, particularly at the low and 
intermediate multipoles of $\ell < 250$ and $\ell \simeq 750$. 
Compared to the baseline model, the DOGE reconstruction reduces excess deviations 
and provides a smoother agreement with the observed data. 
In addition, the $TE$ spectrum shows a noticeable improvement across a wide range 
of multipoles. 
These results indicate that the DOGE framework offers a stable and consistent 
description of both the $TT$ and $TE$ spectra.
As expected, the CPSC-like feature has its strongest impact on large angular 
scales. 
In particular, it leads to improvements at low multipoles of $\ell < 30$, 
where deviations from the baseline model are typically more pronounced. 
Interestingly, this scenario also shows better agreement with the data over 
$\ell \simeq 250$ and around $\ell \simeq 800$, suggesting that the CPSC-like 
feature can influence not only the largest scales but also selected intermediate 
scales.
This behavior is consistent with the localized nature of the reconstructed 
feature and supports its relevance in addressing specific outliers in the 
CMB angular power spectra.
Finally, the power spectrum motivated by the analysis from MRL shows improved 
performance at the largest angular scales due to the presence of a dip in the 
reconstructed primordial scalar power spectrum.
The dip helps the model achieve a better fit to the data over $\ell < 30$,
leading to reduced residuals in this regime. 
However, despite this overall improvement, the GA-generated MRL scenario remains 
unable to fully reproduce the observed behavior at the quadrupole, i.e. at $\ell 
= 2$. 
This suggests that, while the model captures some of the large-scale features in
the data, additional analysis, like the MCMC method, may be required to fit the 
outliers at lowest multipole.

Overall, our analysis demonstrates that different reconstructed features in the
primordial power spectrum affect distinct multipole ranges in the angular power
spectra of the CMB.
The complementary behavior of the DOGE, CPSC-like, and MRL scenarios highlights 
the importance of exploring a variety of reconstruction approaches when studying 
deviations from the standard baseline model.


\section{Impact of varying background parameters}\label{sec:results-cgb}

To assess the sensitivity of the reconstruction of the inflationary power spectrum
using GA to the background cosmological model, we shall now work with modified 
background parameters rather than fixing them at the standard $\Lambda$CDM values. 
The set of background parameters that we shall work with are as 
follows:~$(\Omega_{\mathrm{b}} h^2, \Omega_{\mathrm{c}} h^2,\theta, \tau)
=(0.0225,0.119,1.041,0.05139)$.
These background parameters are the best-fit values obtained from an earlier MCMC 
run with the inflationary power spectrum generated by the DOGE scenario (in this 
context, see Ref.~\cite{Antony:2022ert}). 

\begin{figure*}
\includegraphics[width=0.475\textwidth]{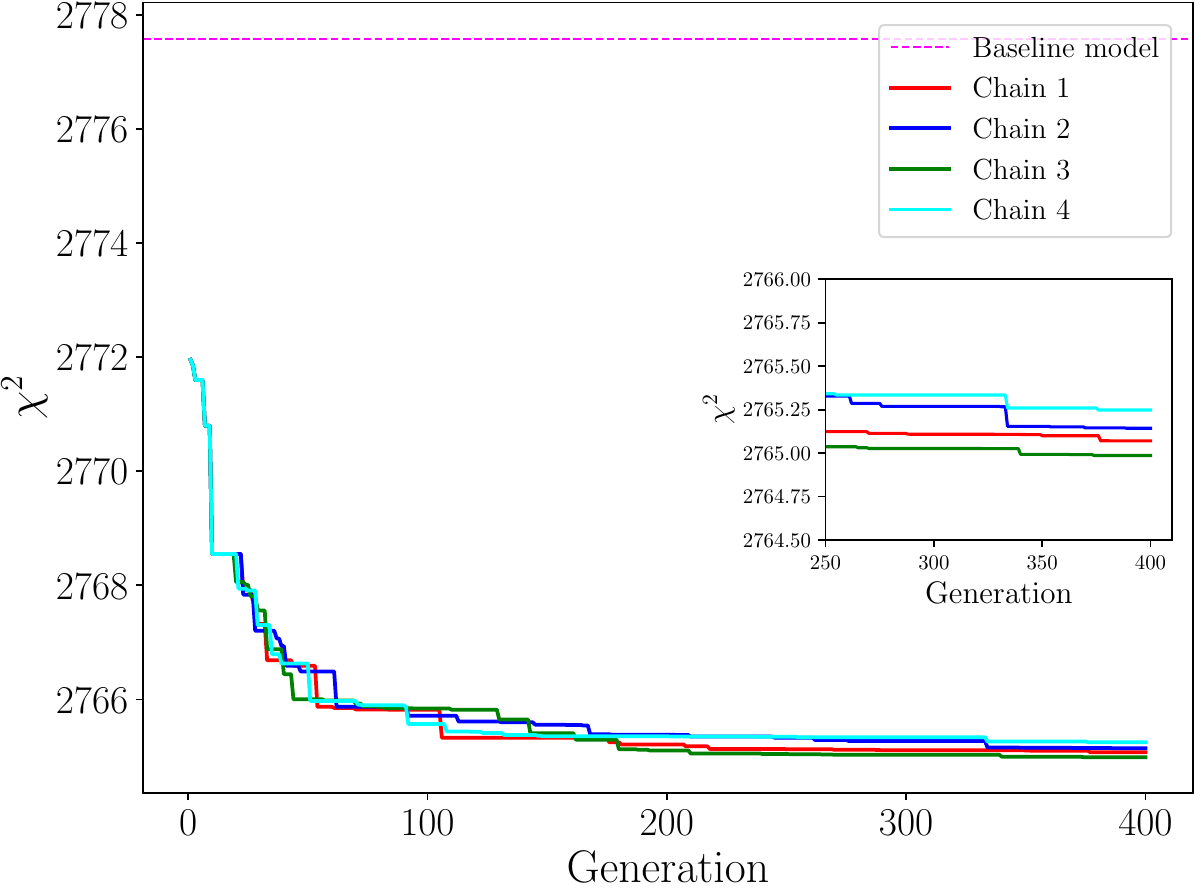}
\hskip 10pt
\includegraphics[width=0.475\textwidth]{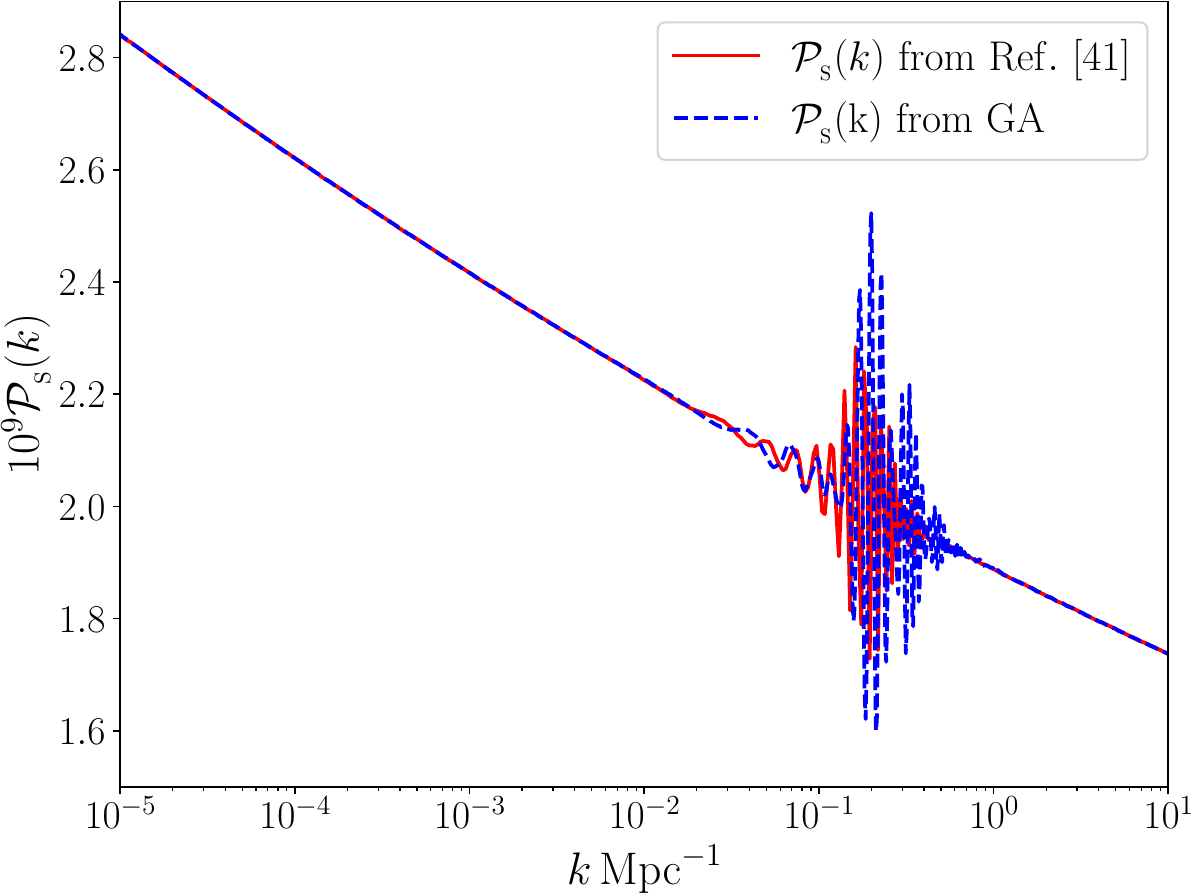}
\caption{We have plotted the evolution of the fitness score $\chi^2$ as a
function of generations in the DOGE~II scenario for four different chains
(on the left).
We have included an inset highlighting the behavior of the fitness score
over the last $150$ generations and we have also indicated the baseline 
fitness score of $\chi^2=2777.59$ (on the left), as in the previous figures.
Note that we have worked with a different set of background parameters than 
the standard $\Lambda$CDM values in arriving at these results.
We have plotted the corresponding scalar power spectrum from the GA 
(on the right, in blue) and we have also included the power spectrum
obtained earlier in this context (on the right, in
red, from Ref.~\cite{Antony:2022ert}).}
\label{fig:doge-II}
\end{figure*}

We too consider a scenario involving DOGE in the first slow roll parameter.
To distinguish from the earlier case, we shall refer to the scenario as DOGE~II. 
We adopt the same grammar template introduced in Eqs.~\eqref{eq:F-test} 
and~\eqref{eq:f-test}. 
The range of priors of the parameters involved is listed in Tab.~\ref{tab:cgb}.
\begin{table}
\centering
\begin{tabular} {|c|c|}
\hline
Parameter &  Prior range\\
\hline
$A$ & $ [0.06, 0.02]   $\\
\hline
$B$ & $[45, 55] $\\
\hline
$C$ & $[15,20] $\\
\hline
$I$ & [0,len(grammar)] \\
\hline
\end{tabular}
\caption{The range of priors for the different parameters that describe 
the function $F(N)$ that modifies the behavior of the baseline first 
slow roll parameter in the DOGE~II scenario [cf. Eq.~\eqref{eq:m-fsrp} 
and Eqs.~\eqref{eq:F-test} and \eqref{eq:f-test}].}\label{tab:cgb}
\end{table}
With the modified background parameters, the GA identifies the following 
$F(N)$ as the best-fit modification function:
\begin{align}
F(N) &= 0.04 \frac{\sin[46.79 (N - 18.22)]}{[1+316.23 (N - 18.22)^2]}\nn \\
&\quad+0.04 \frac{\cos[45.05 (N - 18.40)]}{[1+316.23 (N - 18.40 )^2]}.
\end{align}
The goodness of fit for this reconstruction is $\chi^2 = 2764.98$, which 
corresponds to an improvement of $\Delta \chi^2 = -12.60$ relative to the 
baseline model with modified background parameters. 
In Fig.~\ref{fig:doge-II} we have plotted the evolution of the goodness of
fit value~$\chi^2$ as a function of generations in the same manner as have 
done earlier in Figs.~\ref{fig:test}, \ref{fig:doge-I}, \ref{fig:cpsc} 
and~\ref{fig:mrl}.
In the figure, we have plotted the resulting scalar power spectrum as 
well as the scalar power spectrum obtained in the earlier analysis (see 
Ref.~\cite{Antony:2022ert}). 
Interestingly, we find that, for the above-mentioned choice of background 
parameters and the best-fit scalar power spectrum, the derived values of 
the Hubble constant and the amplitude of the fluctuations in the matter
density turn out to be $H_0 = 67.82 \,\mathrm{km}\,\mathrm{s}^{-1}\,
\mathrm{Mpc}^{-1}$ and $S_8 = 0.821$.
It is interesting to note that these correspond to a $1$-$\sigma$ increase 
in the value of~$H_{0}$ and a $1$-$\sigma$ decrease in $S_8$ compared to 
the predictions of the standard $\Lambda$CDM scenario with a nearly 
scale-invariant scalar power spectrum.

It is important to stress that modifying the background parameters alone, 
i.e. without introducing features in the scalar power spectrum, worsens the 
fit for the baseline model, yielding a goodness of fit of $\chi^2 = 2777.59$. 
This demonstrates that the inclusion of primordial features is essential:~the 
GA-reconstructed oscillatory structures in $\epsilon_1(N)$ provide the necessary
flexibility to reconcile the CMB data with the updated cosmological background. 
Therefore, the improvement in $\chi^2$ arises from a combination of adjustment 
of the background and features in the inflationary scalar power spectrum, 
highlighting the critical role of these features in achieving a better overall fit.

This result indicates that allowing the background parameters to vary may
further enhance the fit to the Planck data.
In other words, certain non-trivial behavior of the first slow roll parameter 
may be more compatible with non-standard background cosmologies.
We should point out that, in this work, we have limited our analysis to the 
Planck 2018 data.
It seems imperative to include the BICEP-Keck 2018~\cite{BICEP:2021xfz}
and the Supernovae, $H_0$, for the Equation of State of Dark Energy (SHOES) 
datasets~\cite{Riess:2021jrx} in the analysis. 
Apart from providing additional constraints on the features in the primordial
scalar power spectrum, incorporating these datasets will also allow us to test 
the robustness of the improvement in the fit to the datasets.
Importantly, such analysis may also offer insights into unexpected correlations 
between primordial physics and late-time cosmology.
This is one of the directions that we are currently working on.


\section{Reconstructing the inflationary potential}\label{sec:potential}

In this section, we shall numerically reconstruct the inflationary potentials 
that lead to the best-fit functional forms of~$\epsilon_1(N)$ in the different 
cases.
From the definition of the first slow roll parameter and the equations of 
motion governing the background, given a form of $\epsilon_{1}(N)$, the 
evolution of the scalar field $\phi(N)$ and the Hubble parameter $H(N)$ can
be expressed as integrals over $e$-folds in the following manner:
\begin{subequations}\label{eq:phi-H}
\begin{align}
\phi(N)=\phi_{\mathrm{i}}-\Mpl\int^{N}_{N_\mathrm{i}} \d N \sqrt{2\epsilon_1(N)},\\
H (N) = H_{\mathrm{i}} \exp \l[ -  \int^{N}_{N_{\mathrm{i}}} \d N \epsilon_{1}(N)\r],
\end{align}
\end{subequations}
where~$\phi_{\mathrm{i}}$ and $H_{\mathrm{i}}$ represent the initial values (at
$N=\Ni$) of the scalar field and the Hubble parameter, respectively. 
In our analysis, we shall set the initial values of these quantities to be 
$\phi_{\mathrm{i}}=6.9\Mpl$ and $H_{\mathrm{i}}= 1.67 \times 10^{-5} \Mpl$.
With the Hubble parameter $H(N)$ and first slow roll parameter $\epsilon_1(N)$ 
at hand, the potential describing the scalar field can be expressed as:
\begin{equation}
V(N)=\Mpl^2 H^2(N)[3-\epsilon_1(N)].\label{eq:vn}
\end{equation}
From the numerical results for $\phi(N)$ and $V(N)$, we can construct the potential 
$V(\phi)$ parametrically. 
Such a reconstruction provides insight into the underlying inflationary potential
that is responsible for generating the best-fit first slow roll parameter. 
\begin{figure}[!t]
\includegraphics[width=0.475\textwidth]{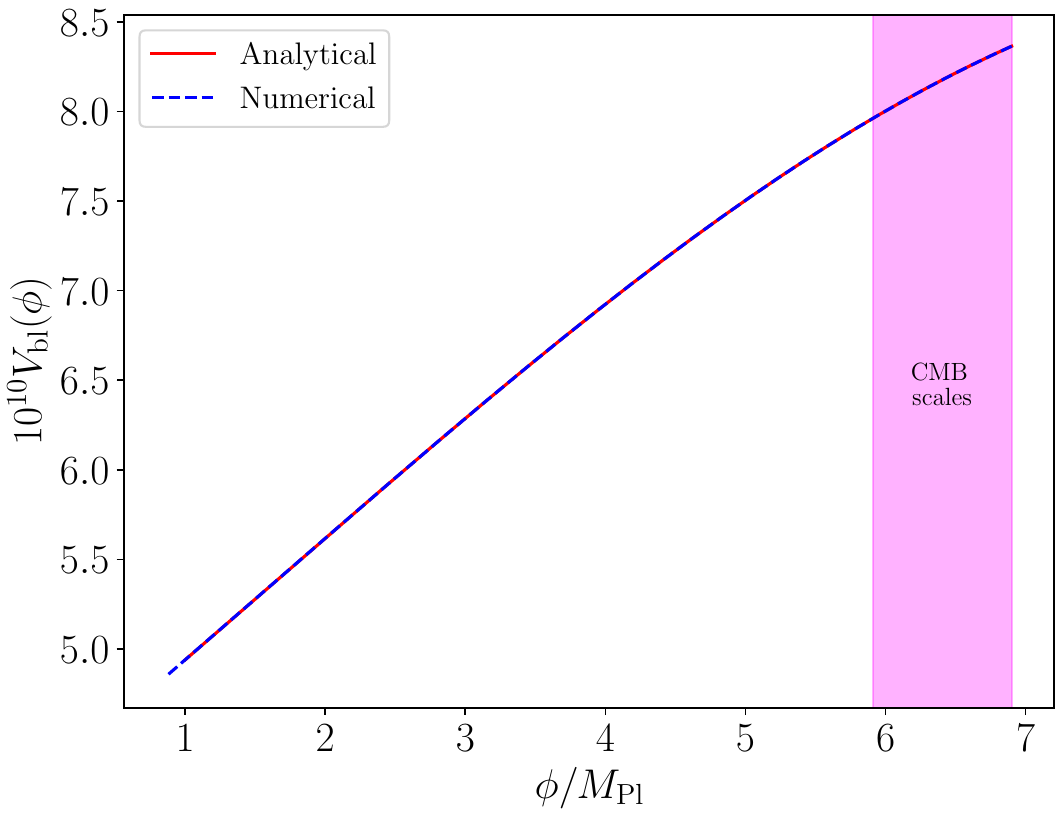}
\caption{We have plotted the analytical expression~\eqref{eq:vbl-phi} as well 
as the corresponding numerical result (in red and blue, respectively) for the 
potential~$V_{\mathrm{bl}}(\phi)$ describing the baseline inflationary model.
We have plotted the potential which leads to $72$ $e$-folds of inflation. 
Also, we have highlighted (in magenta) the range of the field which influences 
the inflationary power spectra over the CMB scales.}\label{fig:vbl}
\end{figure}

\begin{figure*}
\includegraphics[width=0.475\textwidth]{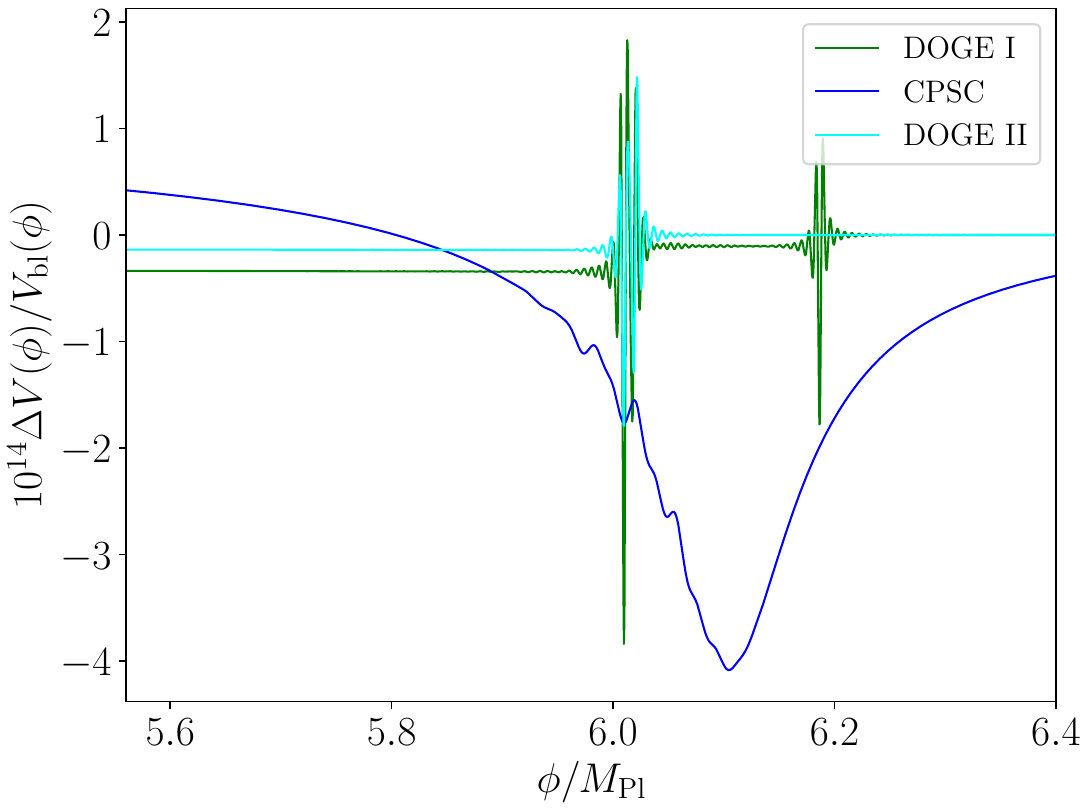}
\includegraphics[width=0.475\textwidth]{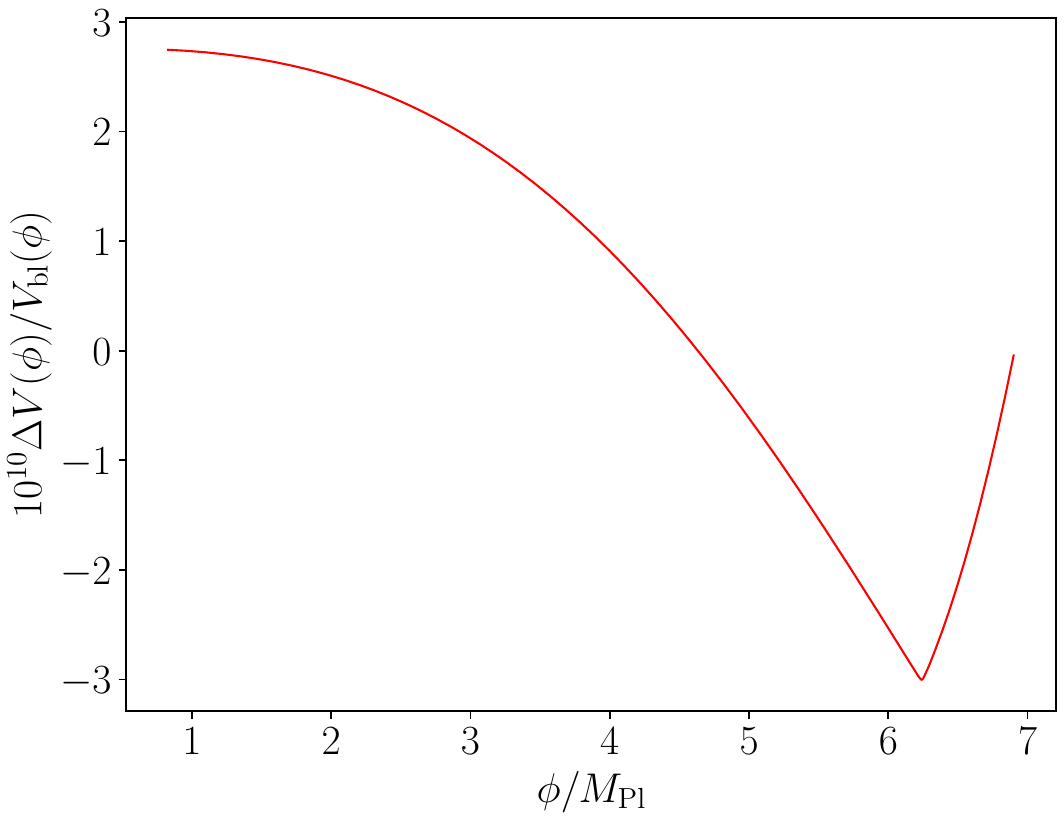}
\caption{We have illustrated the difference between the potential that has been 
determined numerically from the functional forms of the first slow roll parameters 
that lead to the features reconstructed by GA and the potential~\eqref{eq:vbl-phi} 
that describes the baseline model, i.e. $\Delta V(\phi)=V(\phi)-V_{\mathrm{bl}}(\phi)$.
In fact, we have the plotted dimensionless ratio $\Delta V(\phi)/V_{\mathrm{bl}}(\phi)$
for all the cases of interest, viz. DOGE~I, DOGE~II, CPSC (on the left) and MRL 
(on the right).}\label{fig:GA_potential}
\end{figure*}
Before illustrating the reconstructed potential for the different scenarios
leading to features, let us first discuss the inflationary potential describing 
the baseline model.
Interestingly, we find that we can easily arrive at an analytical form for the 
potential describing the baseline model.
On substituting the expression~\eqref{eq:e1-bl} for the functional form for the 
baseline first slow roll parameter~$\epsilon_{1\mathrm{bl}}(N)$ in Eqs.~\eqref{eq:phi-H} 
and carrying out the integrals involved, we obtain $\phi(N)$ and $H(N)$ to be
\begin{subequations}
\begin{align}
\phi(N) &=\phi_{\mathrm{i}}-\f{2}{\epsilon_{2a}}\Mpl 
\l[\sqrt{2\epsilon_{1\mathrm{bl}}(N)}
-\sqrt{2\epsilon_{1\mathrm{bl}}(N_{\mathrm{i}})}\r],\label{eq:phi-N}\\
H(N) &=H_{\mathrm{i}}\exp\l\{-\f{1}{\epsilon_{2a}}\l[\epsilon_{1\mathrm{bl}}(N)
-\epsilon_{1\mathrm{bl}}(N_{\mathrm{i}})\r]\r\}.
\end{align}
\end{subequations}
On substituting these expressions in Eq.~\eqref{eq:vn}, we obtain $V(N)$ to be
\begin{align}
V(N) &=\Mpl^2 H_{\mathrm{i}}^2 \exp\l\{-\f{2}{\epsilon_{2a}}\l[\epsilon_{1\mathrm{bl}}(N)
-\epsilon_{1\mathrm{bl}}(N_{\mathrm{i}})\r]\r\}\nn\\
&\quad\times[3-\epsilon_{1\mathrm{bl}}(N)].
\end{align}
On using the solution~\eqref{eq:phi-N} to express $\epsilon_{1\mathrm{bl}}(N)$
in terms of $\phi$ and substituting the result in the above expression for $V(N)$,
we obtain the potential, say, $V_{\mathrm{bl}}(\phi)$, that describes the baseline
model to be
\begin{align}
V_{\mathrm{bl}}(\phi)
&=\Mpl^2 H_\mathrm{i}^2
\l\{3-\l[\sqrt{\epsilon_{1\mathrm{bl}}(N_{\mathrm{i}})}
-\f{\epsilon_{2a}}{2\sqrt{2}}\f{(\phi-\phi_{\mathrm{i}})}{\Mpl}\r]^2\r\}\nn\\
&\quad\times\exp\l\{\!\f{(\phi-\phi_{\mathrm{i}})}{\Mpl}
\l[\sqrt{\f{\epsilon_{1\mathrm{bl}}(N_{\mathrm{i}})}{2}}
-\f{\epsilon_{2a}}{8} \f{(\phi-\phi_{\mathrm{i}})}{\Mpl}\r]\!\r\}.\label{eq:vbl-phi}
\end{align}
In Fig.~\ref{fig:vbl}, we have plotted the analytical and numerical results for 
the baseline potential $V_{\mathrm{bl}}(\phi)$.

In Fig.~\ref{fig:GA_potential}, we have illustrated the relative change in the 
reconstructed potential (with respect to the potential describing the baseline 
model) that leads to the different features in the inflationary scalar power 
spectrum discussed in the previous section.
Each of these cases highlight how distinct localized features in the inflationary 
potential translate to observable imprints in the scalar power spectrum.
The deviations from the baseline potential provide direct evidence for the manner 
in which the GA pipeline is able to capture non-trivial structures beyond the smooth 
slowly rolling background.
The features in the potential can possibly offer new insights into the dynamics 
that may have occurred during inflation. 
It will be worthwhile to examine the theoretical motivations behind such features
in the potential.


\section{Summary and outlook}\label{sec:summary}

Let us conclude with a brief summary and outlook.

\subsection{Summary}

In this work, we have investigated the possibility that features in the 
primordial scalar power spectrum~$\ps(k)$ can yield a more accurate 
description of the CMB data than the nearly scale-invariant spectrum 
predicted by slow roll inflation. 
To address this, we have introduced a reconstruction pipeline based on the GA,
a machine-learning-inspired optimization framework well-suited to exploring the
parameter spaces.
Our method is designed to reconstruct functional
forms of the first slow roll parameter $\epsilon_1(N)$, which directly controls
the background evolution during inflation and generate localized features in the
scalar power spectrum. 
From this procedure, we identified three distinct effective single-field 
models, each corresponding to an analytical form of $\epsilon_1(N)$
consistent with observational constraints.

The principal strength of our approach lies in its adaptability and 
minimal theoretical bias. 
Unlike conventional reconstruction techniques, which rely on a rigid 
parametrization or pre-specified templates for the inflationary 
potential or the scalar power spectrum, our GA-based framework explores 
the space of inflationary histories in a data-driven manner. 
This allows us to capture a wide variety of functional structures that 
might otherwise remain hidden. 
After a validation through reconstruction of a benchmark scalar power spectrum,
we applied our pipeline to the Planck 2018 data. 

We have reconstructed three distinct features in the primordial scalar power 
spectrum using our GA pipeline. 
In the first case, we demonstrated that introducing a DOGE in the first slow
roll parameter leads to a DOGE in the scalar power spectrum that improves the 
fit to the data. 
In the second case, we successfully reconstructed the power spectrum from a 
popular CPSC model.
Interestingly, using the GA approach, we showed that a scalar power spectrum 
that was originally generated in a two-field model can effectively be 
reconstructed in a single-field model. 
In the third case, we reconstructed features that were entirely data-driven.
To a certain extent, the GA framework allows us to reconstruct such a scalar 
power spectrum in a specific scenario.
Moreover, the resulting reconstructions indicate that GA-generated features in 
the scalar power spectrum lead to a statistically significant improvement in the 
overall likelihood, with $\Delta \chi^2 < -10$ compared to the baseline 
$\Lambda$CDM model with a featureless, nearly scale invariant, power-law primordial
spectrum (in this context, see Tab.~\ref{tab:chi-square}).
\begin{table}
\centering    
\begin{tabular}{|l|c|}
\hline
Model & $\Delta \chi^2$\\
\hline
DOGE~I & $-13.57$ \\
\hline
CPSC & $-13.64$\\
\hline
MRL & $-14.45$\\
\hline
DOGE~II& $-12.60$ \\ 
\hline
\end{tabular}
\caption{The improvement in the goodness of fit with respect to the baseline
model, i.e. $\Delta \chi^2 = \chi^2_{\mathrm{model}}- \chi^2_{\mathrm{baseline}}$,
in the various cases of interest.}\label{tab:chi-square}
\end{table}
Furthermore, when variations in background cosmological parameters are considered,
the GA constructions exhibit different extent of improvement in fit to the data.
This suggest that such models or scenarios may play a role in mitigating persistent 
tensions in the cosmological data, such as those related to the amplitude of the 
lensing potential and the anomalies at the lower multipoles of the CMB.
Importantly, beyond providing a phenomenological description, the reconstructed 
forms of $\epsilon_1(N)$ map onto effective single-field inflationary models 
consistent with the notion of a primordial standard clock, thereby offering a 
physically interpretable inflationary background capable of producing the 
observed CMB features. 
This combination of enhanced statistical performance and theoretical 
consistency underscores the broader utility of our method.

In conclusion, our work establishes that combining machine learning-based 
reconstruction techniques with theoretical inflationary modeling offers a 
robust and versatile framework for exploring the physics of the early 
universe. 
The GA-based pipeline not only enhances the fit to existing cosmological 
data but also provides physically consistent inflationary models with 
well-defined predictive signatures. 
Current and next-generation observatories such as the Atacama Cosmology 
Telescope~(ACT)~\cite{ACT:2025tim}, the South Pole 
Telescope~(SPT) \cite{SPT-3G:2025bzu},
Simons Observatory~(SO)~\cite{SimonsObservatory:2025wwn}, 
Light (Lite) satellite for the studies of B-mode polarization and Inflation 
from cosmic background Radiation Detection (LiteBIRD)~\cite{LiteBIRD:2022cnt}, 
CMB-S4~\cite{CMB-S4:2017uhf, CMB-S4:2016ple, CMB-S4:2020lpa}, and 
Probe of Inflation and Cosmic Origins (PICO) \cite{Alvarez:2019rhd} can,
be expected to constrain (and, possibly detect) the presence of inflationary 
features with unprecedented precision. 
Each of these observatories target complementary regimes of the CMB:~SO 
will probe small scales with high resolution, LiteBIRD will deliver 
cosmic-variance–limited polarization data on large scales, CMB-S4 
promises even finer sensitivity on small angular scales, and PICO aims 
to map the full sky with cosmic-variance–limited accuracy across both 
temperature and polarization. 
In this context, SO and LiteBIRD will be particularly sensitive to the 
suppression in the power on large scales as well as the oscillations 
on small scales.
It would be interesting to investigate whether their combined dataset could
provide decisive statistical evidence about the features in the primordial 
scalar power spectrum.
Thus, the CMB observations over the coming decade may be able to critically 
test the inflationary scenarios we have reconstructed. 
These developments will allow in-depth testing of the GA-reconstructed 
scenarios proposed in this study, potentially uncovering new directions 
of inflationary physics and deepening our understanding of the mechanisms 
that governed the earliest epochs of our universe.


\subsection{Outlook}\label{outlook}

There are a few directions in which the work we have carried out needs to be
extended. 
To begin with, to enrich the reconstructions by GA, we can incorporate different 
datasets in our run. 
Recall that, in this work, we had ignored the contributions due to the tensor 
perturbations. 
If we include the tensors, the data from the BICEP-Keck array can be expected 
to improve the constraints on the tensor-to-scalar ratio on large 
scales~\cite{BICEP:2021xfz}.
Similarly, the data from ACT can be expected to provide improved constraints on
the scalar power spectrum on small scales~\cite{ACT:2025tim}.
The additional datasets can sharpen the ability of the GA to identify features 
in the inflationary scalar power spectrum. 

Secondly, with the templates for the three distinct features now identified,
we can proceed to incorporate the Planck, Bicep-Keck, and ACT datasets and 
perform an MCMC analysis.
In this context, our goal will be to assess whether these features can help 
alleviate, say, the Hubble tension.
We can also conduct a comparative analysis between the GA-identified parameter 
space and the Bayesian analysis of the parameter estimation using MCMC. 

Thirdly, we can extend our analysis to smaller scales, beyond the scales that the 
CMB observations are sensitive to.
Over such scales, the absence of substantial constraints has permitted the exploration
of scenarios that lead to strong departures from slow roll during the later stages
of inflation.
In particular, there has been an interest in examining models that result in enhanced
power on small scales to produce significant number of primordial black holes and
generate scalar-induced, secondary gravitational waves of observable strengths.
Such scenarios have gained importance due to the recent NANOGrav 15-year observations,
which suggest possible evidence for a stochastic gravitational wave 
background~\cite{NANOGrav:2023hde,NANOGrav:2023gor}. 
It has been shown in the literature that secondary, scalar-induced, gravitational 
waves, generated during a phase of reheating which is characterized by a stiffer 
equation-of-state than that of radiation can successfully explain the NANOGrav 
observations with a strong Bayesian evidence~\cite{Maity:2024odg}. 
Using our GA framework, we can reconstruct the functional form of the primordial 
scalar power spectrum to explore the production of primordial black holes and 
the generation of scalar-induced gravitational waves during the epoch of reheating.
Such an analysis will allow us to connect the dynamics during the early and late
stages of inflation as well as reheating.
We are currently working on some of these issues.


\section*{Acknowledgments}

In this work, we have modified and utilized Savvas Nesseris's GA module available 
on GitHub~\cite{Bogdanos:2009ib,Nesseris:2012tt,Arjona:2019fwb,Arjona:2020kco, 
Kamerkar:2022dfu}. 
We wish to thank Akhil Antony, Matteo Braglia and Wuhyun Sohn for the helpful
discussions and for sharing the data files of the primordial scalar power spectrum
for specific models.
DKH and LS would like to thank the Indo-French Centre for the Promotion of 
Advanced Research (IFCPAR/CEFIPRA), New Delhi, India, for support of the proposal
6704-4 titled `Testing flavors of the early universe beyond vanilla models with 
cosmological observations’ under the Collaborative Scientific Research Programme.



\bibliographystyle{apsrev4-1}
\bibliography{references}


\end{document}